\documentclass[pdflatex, sn-nature]{sn-jnl-modified}

\usepackage{graphicx}%
\usepackage{multirow}%
\usepackage{amsmath,amssymb,amsfonts}%
\usepackage{amsthm}%
\usepackage{mathrsfs}%
\usepackage[title]{appendix}%
\usepackage{xcolor}%
\usepackage{textcomp}%
\usepackage{manyfoot}%
\usepackage{booktabs}%
\usepackage{algorithm}%
\usepackage{algorithmicx}%
\usepackage{algpseudocode}%
\usepackage{listings}%
\usepackage{lineno}

\begin{document}

\title[Article Title]{Observation of Quantized Charge Accumulation in a Quantum Anomalous Hall System}

\author[1]{\fnm{Yuanze} \sur{Li}}
\author[1]{\fnm{Jiahao} \sur{Chen}}
\author[2]{\fnm{Renfei} \sur{Wang}}
\author[3]{\fnm{Yifan} \sur{Zhang}}
\author[4]{\fnm{Yingdong} \sur{Deng}}
\author[4]{\fnm{Jin} \sur{Xie}}
\author[3,5]{\fnm{Xufeng} \sur{Kou}}
\author[2]{\fnm{Yang} \sur{Liu}}
\author*[1,6]{\fnm{Tian} \sur{Liang}}\email{tliang@mail.tsinghua.edu.cn}

\affil[1]{\orgname{State Key Laboratory of Low Dimensional Quantum Physics, Department of Physics, Tsinghua University}, \orgaddress{\city{Beijing} \postcode{100084}, \country{People's Republic of China}}}

\affil[2]{\orgname{International Center for Quantum Materials, Peking University}, \orgaddress{\city{Beijing} \postcode{100871}, \country{People's Republic of China}}}

\affil[3]{\orgname{School of Information Science and Technology, ShanghaiTech University}, \orgaddress{\city{Shanghai} \postcode{201210}, \country{People's Republic of China}}}

\affil[4]{\orgname{School of Physical Science and Technology, ShanghaiTech University}, \orgaddress{\city{Shanghai} \postcode{201210}, \country{People's Republic of China}}}

\affil[5]{\orgname{ShanghaiTech Laboratory for Topological Physics, School of Physical Science and Technology, ShanghaiTech University}, \orgaddress{\city{Shanghai} \postcode{201210}, \country{People's Republic of China}}}

\affil[6]{\orgname{Frontier Science Center for Quantum Information}, \orgaddress{\city{Beijing} \postcode{100084}, \country{People's Republic of China}}}

\abstract{The quantum anomalous Hall effect in magnetically doped topological insulators exhibits a quantized Hall conductance $\sigma_{xy} = e^2/h$ arising from the two-dimensional surface states. While conventional transport probes confirm this quantization, they remain insensitive to the field-induced surface charge accumulation as a direct manifestation of $\sigma_{xy}$. Here, we experimentally validate an out-of-plane capacitive method that directly detects this quantized charge accumulation in a quantum anomalous Hall system. Using Corbino and simple disk devices, we measure charge accumulation proportional to field variation $\Delta B$, with dissipation characterized by longitudinal conductance $\sigma_{xx}$ and frequency $f$. A quantitative dissipation model extracts the intrinsic quantized charge density $\eta_0 = (e^2/h)\Delta B$, which is confirmed through $f$- and $\sigma_{xx}$-dependent measurements. Under ultra-low dissipation ($\sigma_{xx} \approx 10^{-9}$ S), we directly resolve the fully quantized charge accumulation. This methodology establishes a direct charge-accumulation probe and provides a pathway toward detecting the topological magnetoelectric effect, a condensed matter manifestation of the four-dimensional quantum Hall effect.
}

\maketitle

\section*{Introduction}

The quantum anomalous Hall (QAH) effect, arising as the two-dimensional (2D) surface state of a three-dimensional topological insulator (3D TI), represents a hallmark phenomenon in topological materials, attracting sustained attention since its theoretical prediction and first experimental realization \cite{yu2010QAH, kane2010TI, SCZhang2011, ChangCZ2013_QAH_CBST}. The QAH system features a quantized Hall conductance and vanishing longitudinal conductance similar to the quantum Hall (QH) effect. However, the QAH effect requires no external magnetic field, as it arises from the intrinsic magnetization of the material. Magnetically ordered 3D TIs serve as the primary material platform for realizing the QAH effect \cite{ChangCZ2013_QAH_CBST, KouXufeng2014_QAH_CBST, Tokura2014_QAH_CBST, ZhangYuanbo2020_QAH_MBT}, where the top and bottom surfaces of a thin film sample each contribute a half-quantized Hall conductance, $ \sigma^\mathrm{s}_{xy} = \pm e^2/2h $, with the sign depending on the surface magnetization direction. When combined, both surfaces yield the integer-quantized Hall conductance $\sigma_{xy} = \pm e^2/h$. The half-quantization of the surface state originates from a massive Dirac Hamiltonian, which arises when the exchange interaction from surface ferromagnetism gaps the surface states of 3D TIs \cite{qi2008topological, morimoto2015topological}. The analysis based on Berry curvature yields $\pm e^2/2h$ for each surface, establishing a topologically nontrivial surface state.

The remarkable signatures of the QAH effect—quantized Hall conductance and vanishing longitudinal conductance observed in Hall-bar experiments—are conventionally explained by the presence of a one-dimensional (1D) chiral edge state \cite{qi2008topological, zhang2013chiral}. The edge channel forms as a domain wall where the surface Hall conductance changes from $+e^2/2h$ to $-e^2/2h$, yielding a chiral edge state at the physical boundary of the sample. However, this model oversimplifies a fundamentally 2D effect into a boundary phenomenon. The QAH state originates from the quantized Hall conductance of the entire surface, in other words, it is intrinsically a 2D phenomenon. Consequently, although transport measurements in Hall-bar geometries are technically convenient and powerful for confirming quantization, they can only provide an indirect probe of the underlying 2D surface properties as the interpretation of the Hall signal in Hall-bar geometry is unnecessarily convoluted by the edge state. Therefore, developing experimental techniques that can directly access the intrinsic surface Hall conductance, independent of sample edges, is essential for a more complete understanding of the QAH effect.

To this end, a few experimental approaches have been developed. One prominent method is the optical Hall measurement based on time-domain terahertz spectroscopy (TDTS), which detects the Faraday and Kerr rotations as signatures of the surface Hall conductance, while avoiding contributions from edge channels by spatially isolating the beam spot from the sample boundary \cite{WuLiang2016_3DTI_THz_optical, Tokura2016_QAH_THz_optical,Shimano2010_QHE_THz_optical,Shimano2013_graphene_THz_optical}. Another approach, inspired by Laughlin's charge pumping model \cite{laughlin1981}, measures the quantized adiabatic charge transfer between the inner and outer contacts of a Corbino disk sample under field variation \cite{1992_Russia_GaAs_chargepumping&Streda, RIKEN_CBST_chargepumping}, yielding quantized Hall conductance $\sigma_{xy}$. In the Corbino disk geometry, the charge pumping proceeds entirely through the adiabatic variation of 2D states, eliminating the involvement of edge channels.

While the above-mentioned approaches \cite{WuLiang2016_3DTI_THz_optical, Tokura2016_QAH_THz_optical, Shimano2010_QHE_THz_optical, Shimano2013_graphene_THz_optical, 1992_Russia_GaAs_chargepumping&Streda, RIKEN_CBST_chargepumping} successfully detect the two-dimensional quantum Hall effect (2D QHE) without any involvement of 1D edge channels, the techniques rely on in-plane geometries and are thus accessible only to 2D QHE. In this work, we develop an out-of-plane capacitive method for measuring charge accumulation, which is accessible to both 2D QHE and the four-dimensional quantum Hall effect (4D QHE) \cite{SCZhang2001_4DQH, qi2008topological, Vanderbilt2009_3DTI} (see the Supplementary Information Section~VI for details), and validate it by measuring the field-induced net charge accumulation in QAH systems (a 2D QH system).

\section*{Results}

\subsection*{Principles and experimental setups}

\begin{figure}[h!]
  \centering
  \includegraphics[width=1\linewidth]{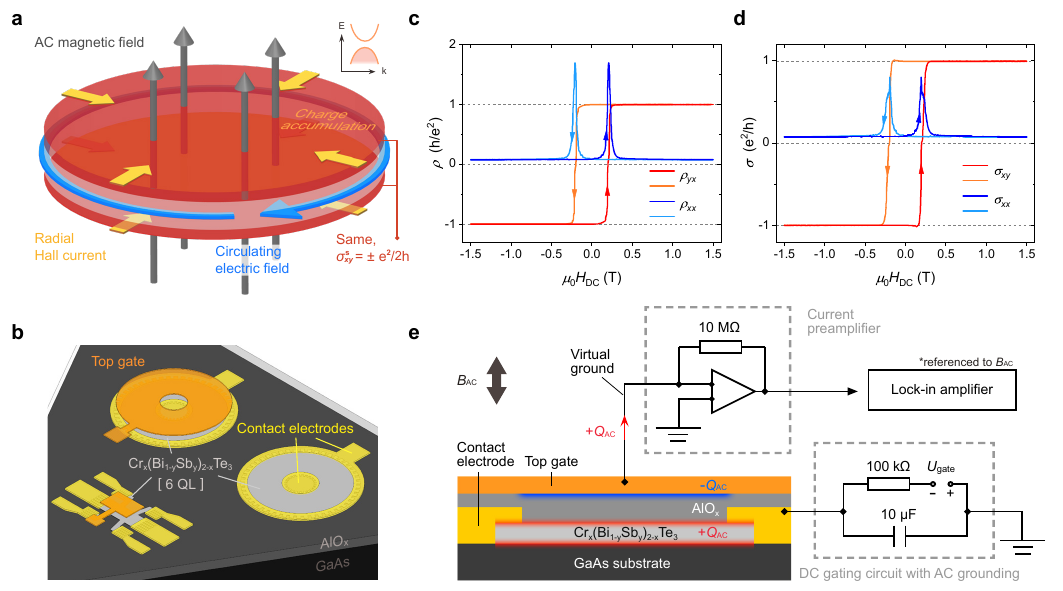}
  \caption{\textbf{Principles of charge accumulation, basic transport of 6-QL CBST sample, and measurement configuration.} 
  \textbf{a}, Schematic of charge accumulation in a thin-film QAH sample. The top and bottom surfaces each contribute $\sigma^\mathrm{s}_{xy}=\pm e^2/2h$ due to the magnetically gapped 2D surface state consisting of a massive Dirac cone. A time-varying out-of-plane field $B_\mathrm{AC}$ (vertical arrows) induces a circulating Faraday electric field (blue arrow) that drives a radial Hall current (yellow arrows) and builds surface charge (red shading) on both surfaces. 
  \textbf{b}, Device layout including a Hall bar and two Corbino disks on a GaAs(111) substrate. The gray and yellow areas underneath the transparent AlO$_x$ dielectric indicate the 6-QL CBST films and contact electrodes, respectively. Top gates are shown in orange.  The top gate of the right Corbino disk is omitted to reveal the underlying structures.
  \textbf{c,d}, DC-field dependence of resistivity ($\rho_{xx},\rho_{yx}$) (\textbf{c}) and the corresponding conductance ($\sigma_{xx},\sigma_{xy}$) (\textbf{d}) measured on the Hall bar at $T\approx200$ mK and zero gate voltage. Data for upward and downward DC field sweeps are labeled with arrows.
  \textbf{e}, Schematic configuration of the charge accumulation measurement. A cross-section of the Corbino disk structure is displayed with the measurement circuits. Red and blue luminescence indicates the AC charge accumulation $Q_\mathrm{AC}$ in the QAH sample and the induced opposite charge $-Q_\mathrm{AC}$ under the top gate, respectively, generated by $B_\mathrm{AC}$ perpendicular to the sample plane. (The charge accumulation value $Q_\mathrm{AC}$ does not include the possible charge in the sample region beneath contact electrodes, where the charge accumulation is shielded.) The circuit on the top describes the charge measurement, where $Q_\mathrm{AC}$ is measured by a current amplifier at virtual ground and then recorded by a lock-in amplifier referenced to $B_\mathrm{AC}$. The circuit on the right provides a DC gating voltage, while retaining the function of an effective AC short circuit due to the large parallel capacitance.}
  \label{fig:fig1}
\end{figure}

A clear physical picture of the intrinsic charge accumulation emerges from the Hall response of a QAH system under a time-varying magnetic field. For conceptual clarity, we illustrate this using a simple disk geometry as shown in Fig.~\ref{fig:fig1}\textbf{a}, but it is important to note that the resulting charge accumulation is a general 2D property and thus independent of the sample's shape. When the time-varying field is applied perpendicular to the sample plane, Faraday's law dictates that an in-plane circulating electric field is induced on both surfaces. This electric field, in turn, drives a radial Hall current on each surface due to its half-quantized Hall conductance ($\sigma^\mathrm{s}_{xy} = \pm e^2/2h$). For a single surface, the total outward current is given by:
\begin{equation}
I_{\mathrm{out}} = \sigma^\mathrm{s}_{xy} \oint_{\partial S} \mathbf{E} \cdot d\mathbf{l} = -\sigma^\mathrm{s}_{xy} \frac{\partial \Phi}{\partial t},
\end{equation}
Here $\Phi \equiv \int_S B_z dS$ is the magnetic flux through an area $S$ enclosed by the contour $\partial S$, and $B_z$ is the magnetic field component perpendicular to the sample plane. Assuming negligible dissipation, this radial Hall current leads to a net accumulation of surface charge. Since the current flows only during the field's variation, the resulting charge density is proportional to the change in magnetic field, given by 
\begin{equation}
  \eta_0 = \sigma_{xy} \Delta B
\end{equation}

where $\sigma_{xy}$ is the total Hall conductance of both surfaces combined, and $\Delta B$ is the change in the magnetic field perpendicular to the sample plane. A general derivation of $\eta_0$ is provided in Supplementary Information Section~VII. This model provides a direct physical link between the accumulated surface charge and the surface Hall conductance, which serves as the key observable in our experiment. Furthermore, in this ideal case without dissipation, detection of charge accumulation $\eta_0 = \sigma_{xy} \Delta B$ can serve as a highly sensitive probe of $\Delta B$, i.e., essentially magnetization. This is another way of viewing the conceptual difference between the charge accumulation measurement and the Hall conductance measurement.

Once the dissipation is taken into account, the measured charge accumulation is given by $Q_\mathrm{a.} = g A_\mathrm{S} \eta_0$, where $\eta_0$ is the intrinsic and quantized 2D charge density related to the Hall conductance $\sigma_{xy}$ obtained above, and $g$ is a dissipation factor that depends on the longitudinal conductance $\sigma_{xx}$ and the measurement frequency $f$. Therefore, detecting charge accumulation differs fundamentally from measuring $\sigma_{xy}$ alone. While $\sigma_{xy}$ remains quantized as long as $\sigma_{xx}$ is sufficiently small, the charge accumulation is generally significantly affected by the dissipation factor $g(\sigma_{xx}/fC_{\mathrm{TG}})$, which is a function of the ratio $\sigma_{xx}/fC_{\mathrm{TG}}$ ($C_\mathrm{TG}$ is the geometric capacitance of the top gate dielectric) and is related to the sample geometry. Crucially, even for a small $\sigma_{xx}$ that does not affect the quantization of the Hall conductance $\sigma_{xy}$, the factor $g(\sigma_{xx}/fC_{\mathrm{TG}})$ can deviate substantially from unity. In the following paragraphs, we develop a systematic analytical model for the general case where $g(\sigma_{xx}/fC_{\mathrm{TG}})\neq 1$. This model enables the extraction of the quantized intrinsic charge accumulation from experimental data. This general treatment corresponds to actual experimental conditions encountered when working with realistic QAH thin film samples. Here, we emphasize that our model describes a non-equilibrium state, in which charge accumulation and dissipation arise as responses to electric fields induced by time-dependent alternating-current (AC) magnetic fields. This stands in sharp contrast to the Streda formula \cite{streda1982}, which also relates charge accumulation and magnetic field with a quantized coefficient, but applies strictly to an equilibrium regime. For further details, we refer the reader to the Supplementary Information Section~XII.

In the actual experiment, a six-quintuple-layer (6-QL) chromium-doped 3D TI $\mathrm{(Bi,Sb)}_2\mathrm{Te}_3$ (CBST) sample was prepared for the measurement, which was grown on a GaAs substrate by molecular beam epitaxy (MBE), as described in previous studies \cite{KouXufeng2022-CBST-thickness}. As shown in Fig.~\ref{fig:fig1}\textbf{b}, a single chip of the sample was fabricated into a device including both Hall-bar and Corbino disk structures, each with an independent top gate. The Corbino disk structure is used for the primary measurement of field-induced charge accumulation. Its geometry with inner and outer contacts allows for the simultaneous measurement of charge pumping signals and the $\sigma_{xx}$ via two-terminal measurements, providing crucial in-situ information unavailable in a simple disk-shaped sample. The Hall bar structure serves as a reference for conventional transport measurement, where the measured Hall and longitudinal resistivity ($\rho_{yx}$ and $\rho_{xx}$) relative to the magnetic field are shown in Fig.~\ref{fig:fig1}\textbf{c}, and the corresponding Hall and longitudinal conductance ($\sigma_{xy}$ and $\sigma_{xx}$) are shown in Fig.~\ref{fig:fig1}\textbf{d}. The presence of hysteresis loops with quantized $\sigma_{xy} = \pm e^2/h$ at zero field shows a clear signature of the QAH effect. These data were measured at the temperature of approximately 200 mK with zero top gate voltage. Due to the inhomogeneity of the sample, the Fermi energy of the Corbino disk region is slightly different from that of the Hall-bar region, as concluded from the optimal gate voltages that minimize the longitudinal conductance, which are compared in the Supplementary Information Fig.~S1. All subsequent measurements on the Corbino disk sample were performed at its optimal gate voltage, which is defined as the voltage difference from the top gate to contact electrodes. 

The 2D charge accumulation is measured via the top gate, which capacitively couples to the sample, as schematically shown in Fig.~\ref{fig:fig1}\textbf{e}. A small AC magnetic field (superimposed on the direct-current (DC) field) is applied perpendicular to the sample plane, which induces the target charge accumulation. The top gate is connected to a virtual ground through a current preamplifier. This configuration allows the charge accumulation in the sample to induce an image charge with the opposite value on the gate, while the charge variation on the gate is probed by the current preamplifier. Therefore, the charge accumulation $Q_\mathrm{a.}$ in the sample is transferred as a current signal $I = 2 \pi f Q_\mathrm{a.} i$, where $f$ is the frequency of the AC magnetic field and $i$ represents a quadrature phase shift. The intrinsic charge accumulation signal is calculated, in turn, based on the measured current signal. In addition, a specially designed DC voltage circuit is connected to the contact electrodes, which enables DC gating of the sample while providing effective grounding in AC circuits without influencing AC signals (see Methods for details).

While the charge accumulation originates from quantized Hall conductance, it is simultaneously subject to dissipation from the sample's finite longitudinal conductance. The accumulated charge $Q_\mathrm{a.}$ itself creates an in-plane Coulomb potential difference that drives a dissipative radial current, causing the charge to decay. The nearby grounded top gate effectively screens this potential, but a residual potential proportional to $\eta A_\mathrm{S} / C_\mathrm{TG}$ (where $\eta$ is the accumulated charge density, $A_\mathrm{S}$ is the sample area, and $C_\mathrm{TG}$ is the gate-to-sample capacitance) persists. This results in a finite decay of charge accumulation in our experiments, which modifies the ideal quantized response. However, this dissipation can be systematically characterized, allowing us to extract the intrinsic, unattenuated charge accumulation value from experimental signals, which is analyzed in subsequent sections.

\subsection*{Charge accumulation and charge pumping in QAH samples}

\begin{figure}[h!]
  \centering
  \includegraphics[width=1\linewidth]{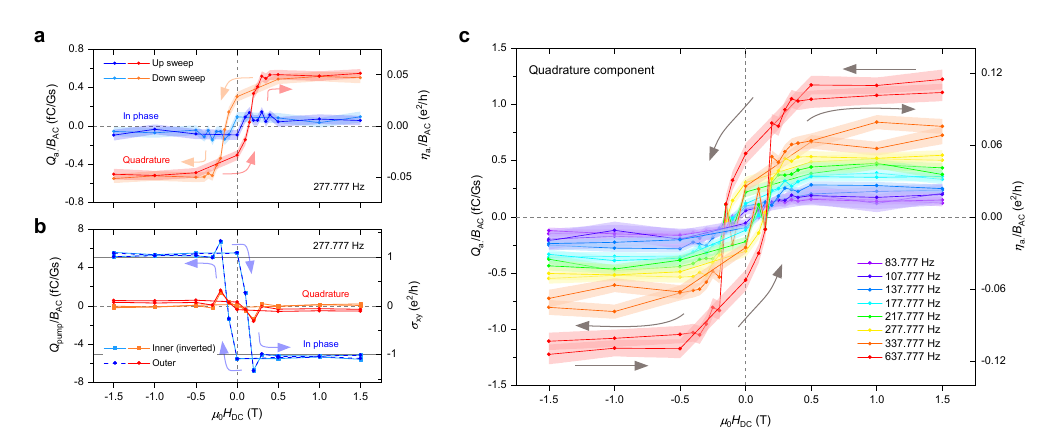}
  \caption{\textbf{Quantized charge pumping and frequency-dependent charge accumulation in the QAH Corbino disk sample.} 
  \textbf{a}, DC field dependence of charge accumulation at $f = 277.777$ Hz. The left axis shows the total accumulated charge $Q_\mathrm{a.}$ normalized by the AC magnetic field $B_\mathrm{AC}$, while the right axis shows the corresponding normalized charge density (spatial average) $\eta_\mathrm{a.}/B_\mathrm{AC}$ in units of $e^2/h$. In-phase and quadrature components are plotted separately. Shaded bands denote standard errors, which are calculated from repeated measurements. Data for upward and downward DC field sweeps are distinguished by color and labeled with arrows. 
  \textbf{b}, DC field dependence of normalized charge pumping $Q_\mathrm{pump}/B_\mathrm{AC}$ (charge flowing out of the sample) measured at $f = 277.777$ Hz (the same conditions as in \textbf{a}). In-phase and quadrature components are displayed. Data measured at inner and outer contacts of the Corbino disk are plotted together. The inner-contact trace is inverted for direct comparison with the outer-contact trace. The arrows accompanying the in-phase data label the sweep directions of the DC field. The right axis shows the calculated Hall conductance $\sigma_{xy}$ corresponding to the in-phase charge signal $Q_\mathrm{pump}^\mathrm{in-phase}$. 
  \textbf{c}, Frequency dependence of the quadrature component of $Q_\mathrm{a.}/B_\mathrm{AC}$. DC field-dependent charge-accumulation data measured at frequencies from $83.777$ Hz to $637.777$ Hz are shown. Axes match \textbf{a}. Upward and downward sweeps are distinguished by arrows. Shaded bands denote standard errors.}
  \label{fig:fig2}
\end{figure}

The charge accumulation signals measured under a 277.777 Hz AC magnetic field at approximately 200 mK are presented in Fig.~\ref{fig:fig2}\textbf{a}, displaying both the in-phase (real) and quadrature (imaginary) components. All subsequent data in this paper, except for longitudinal conductivity dependence in Fig.~\ref{fig:fig3}\textbf{c}, were measured at this sample temperature. To remove background crosstalk, all signals have been antisymmetrized with respect to the DC magnetic field for extracting genuine signals (see the Supplementary Information Section~II for details). The data exhibit clear hysteresis loops with plateaus at high DC fields, which are consistent with the Hall conductance characteristics of the QAH effect. To facilitate comparison with the expected quantization, the total charge signal is normalized by the sample area of 2.757 $\mathrm{mm}^2$ (average accumulated charge density $\eta_\mathrm{a.}$) and presented in units of $e^2/h$ on the right vertical axis. The AC magnitude of the signal is much smaller than the quantized value, while the phase of the signal is nearly in quadrature with the AC field, which is consistent with an RC (resistance-capacitance) decaying situation as discussed before (resistance refers to $1/\sigma_{xx}$, and capacitance refers to $C_\mathrm{TG}$). Notably, while the Hall conductance ought to remain quantized near zero DC field, the charge accumulation signal decays more severely. This behavior is a direct consequence of the enhanced dissipation, as the longitudinal conductance increases near zero field in Fig.~\ref{fig:fig1}\textbf{c}.

For comparison, Laughlin charge pumping signals measured under the same AC field conditions are shown in Fig.~\ref{fig:fig2}\textbf{b} (see the Supplementary Information Fig.~S3 for detailed methods and additional data, where the idea draws on previous studies \cite{RIKEN_CBST_chargepumping}). The right vertical axis in Fig.~\ref{fig:fig2}\textbf{b} represents the Hall conductance corresponding to the charge pumping signal calculated by $\sigma_{xy} = \frac{2 \ln{(r_2/r_1)}}{\pi (r_2^2-r_1^2)} \frac{Q_\mathrm{pump}^\mathrm{in-phase}}{B_\mathrm{AC}}$ ($r_1$ and $r_2$ refer to the inner and outer ring radii of the Corbino disk sample), which reveals a full quantization of in-phase signals. The transition field range for the conductance reversal (0.1 T to 0.2 T) aligns perfectly with the charge accumulation data. In particular, the pumped charges at the inner and outer contacts of the Corbino disk sample show a slight difference, which reflects the charge accumulation according to charge conservation. While this difference offers an alternative way to measure charge accumulation, its precision is inevitably lower than the direct gate method (see the Supplementary Information Fig.~S4 for details).

Although the charge accumulation and the Laughlin charge pumping both stem from Hall conductivity ($\sigma_{xy}$), they represent fundamentally different phenomena. Charge pumping describes the transport of charge across the sample and does not necessarily require the accumulation of charge. In particular, the inner and outer charge pumping signals can take (almost) the same value as described by Fig.~\ref{fig:fig2}\textbf{b}. By contrast, charge accumulation is a buildup of charge density itself on the 2D surface. This accumulated charge generates an in-plane electrostatic field $E_r$, which is precisely why the signal is susceptible to dissipation via $\sigma_{xx}$ and is also what makes it a powerful probe for distinguishing the system's states, whose information cannot be obtained by simply measuring the Hall conductivity $\sigma_{xy}$ alone.

To systematically investigate the role of dissipation, we measured charge accumulation signals across a range of frequencies, as shown in Fig.~\ref{fig:fig2}\textbf{c}. The quadrature component is plotted, as it contains the primary signal. To isolate frequency as the key variable, we adjusted the AC field amplitude at each frequency (balancing eddy current heating) to maintain a constant sample temperature, thereby keeping the longitudinal conductance fixed under the same DC field (see the Supplementary Information Fig.~S5 for calibration). The data in Fig.~\ref{fig:fig2}\textbf{c} show a consistent dependence on the DC field at all frequencies, exhibiting clear plateaus at high fields. Crucially, the charge signal amplitudes increase with frequency. This demonstrates that the detrimental effects of dissipation are mitigated at faster measurement timescales. Although the experimental setup currently limits our maximum frequency, this trend suggests that the fully quantized charge accumulation signal could eventually be reached under sufficiently suppressed decay.

\subsection*{Dissipation model and the intrinsic quantized charge accumulation}

\begin{figure}[h!]
  \centering
  \includegraphics[width=1\linewidth]{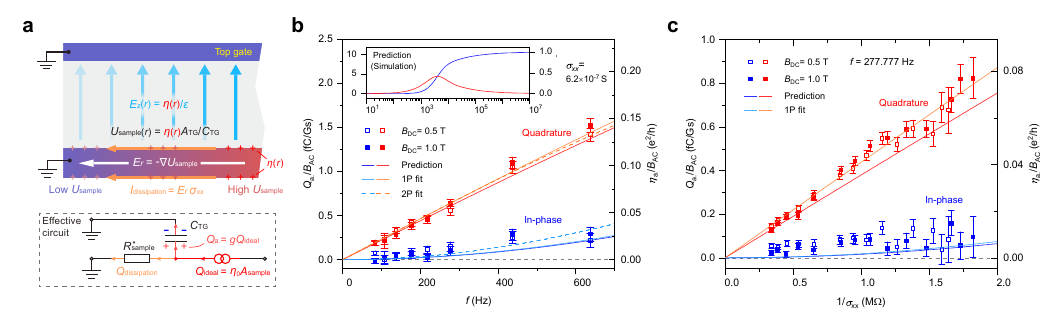}
  \caption{\textbf{Dissipation model and quantitative fits for the field-induced charge accumulation.} \textbf{a}, Schematic of the dissipation model (top) and effective circuit (bottom). The three layers from top to bottom in the schematic are the grounded top gate, the AlO$_x$ dielectric, and the CBST film (connected to a grounded contact at left). Vertical arrows of varying shade represent the out-of-plane electric field $E_z$ in the dielectric, which is proportional to the charge density $\eta$ in the sample. This out-of-plane electric field determines the potential $U_\mathrm{sample}$ in the sample and the in-plane electric field $E_r$, which generates the dissipative current $I_\mathrm{dissipation}$. The effective circuit qualitatively summarizes the dissipation mechanism into three elements: a dissipative resistance $R_\mathrm{sample} \propto 1/\sigma_{xx}$, the gate-to-sample capacitance $C_\mathrm{TG}$, and a current source that pumps the intrinsic charge accumulation $Q_\mathrm{ideal} = \eta_0 A_\mathrm{S}$ into the sample. The measured charge accumulation is $Q_\mathrm{a.} =g(\sigma_{xx}/fC_\mathrm{TG})\,Q_{\mathrm{ideal}}$.
  \textbf{b,c}, Frequency ($f$) and longitudinal conductance ($\sigma_{xx}$) dependence of the charge accumulation and parameter fits. The data are measured at fixed sample longitudinal conductance of $6.2 \times 10^{-7}$ S (\textbf{b}) and fixed frequency of $277.777$ Hz (\textbf{c}), respectively. These fixed parameters are used in the simulations and fits. The left and right vertical axes show the total accumulated charge $Q_\mathrm{a.}$ and the corresponding average charge density $\eta_\mathrm{a.}$, respectively, both normalized by the AC magnetic field. In \textbf{c}, the horizontal axis is plotted as $1/\sigma_{xx}$ for clarity. In-phase (blue) and quadrature (red) components of the data are displayed separately. The error bars on data points represent standard errors calculated from repeated measurements. Solid lines are the prediction of charge accumulation signals using the quantized value ($\eta_0^{(\mathrm{unit} B)} = e^2/h$) and Eq. \eqref{accumulated_charge_Corbino}. The inset in \textbf{b} uses the same axes as \textbf{b} and shows an extended frequency range of the prediction, illustrating the crossover from dissipation-dominated to near-ideal charge accumulation. Light solid lines are one-parameter fits of Eq. \eqref{accumulated_charge_Corbino} with free $\eta_0^{(\mathrm{unit} B)}$. Light dashed lines in \textbf{b} are two-parameter fits of Eq. \eqref{accumulated_charge_Corbino} with free $\eta_0^{(\mathrm{unit} B)}$ and dissipation parameter $\lambda_f = \frac{ \sigma_{xx} A_\mathrm{TG}}{C_\mathrm{TG}}$. The fits were performed on the combined data from both DC fields.}
  \label{fig:fig3}
\end{figure}

Since dissipation modifies the quantized signal, we employ a quantitative model to extract the intrinsic, undissipated value of the charge accumulation. A conceptual schematic of this model is shown in Fig.~\ref{fig:fig3}\textbf{a}. The charge accumulation density $\eta$ determines the potential $U_\mathrm{sample}$ in the Corbino disk sample, which generates the dissipation current $I_\mathrm{dissipation}$ and alters the value of the accumulated charge. A qualitative circuit is displayed below the schematic, where the microscopic dissipation mechanism is simplified as an effective resistance $R_\mathrm{sample}$. The dissipative charge $Q_\mathrm{dissipation}$ flowing through this resistance divides the ideal charge accumulation $Q_\mathrm{ideal} = \eta_0 A_\mathrm{S}$, leaving a real accumulated charge as $Q_\mathrm{a.} = g Q_\mathrm{ideal}$ in the sample. In a quantitative model, three factors govern the dissipation: the gate-to-sample capacitance ($C_\mathrm{TG}$), which sets the potential rise generated per unit of accumulated charge; the longitudinal conductance ($\sigma_{xx}$), which determines the dissipative current generated by the potential rise; and the AC frequency ($f$), which dictates the timescale for the decay. As derived in the Supplementary Information Section~VII, the physical picture of charge accumulation with dissipation is captured by a differential equation for the complex charge density, $\tilde{\eta}$ (the complex number corresponds to the phase information of the AC signal):

\begin{equation}
    (1 + i\frac{ \sigma_{xx} A_\mathrm{TG}}{2 \pi f C_\mathrm{TG}} \nabla^2) \tilde{\eta} = \sigma_{xy} B_{\mathrm{AC}} = \eta_0^{(\mathrm{unit} B)} B_{\mathrm{AC}}
    \label{distribution_differential_equation}
\end{equation}

where $A_\mathrm{TG}$ is the top gate area (note that $A_\mathrm{TG}$ is slightly larger than the sample area because the top gate also covers part of the contact electrodes; also, $C_\mathrm{TG}$ defined here includes the capacitance of the overlapping top gate and contact area). The analytical solution to this equation for our Corbino disk geometry (with inner ring radius of $r_1$ and outer ring radius of $r_2$) provides a testable model, which corresponds to a total accumulated charge $\tilde{Q}_\mathrm{a.}$ as

\begin{equation}
  \tilde{Q}_\mathrm{a.} / B_{\mathrm{AC}} = \int_{r_1}^{r_2} 2 \pi r \, dr \, \tilde{\eta}(r) =  g(\sigma_{xx}/fC_\mathrm{TG}) \pi (r_2^2 - r_1^2) \eta_0^{(\mathrm{unit} B)}
\label{accumulated_charge_Corbino}  
\end{equation}

where $\pi (r_2^2 - r_1^2)$ is the area of the Corbino disk sample, and $g(\sigma_{xx} /fC_\mathrm{TG})$ is the dissipation factor discussed previously. Its detailed expression is

\begin{equation}
\begin{split}
  g(\sigma_{xx}/fC_\mathrm{TG}) = \, &\Big\{ 1 + \frac{2}{k (r_2^2 - r_1^2) \left(J_0(kr_1)Y_0(kr_2) - J_0(kr_2)Y_0(kr_1)\right)} \times \\
  &[(Y_0(kr_1) - Y_0(kr_2)) \cdot (r_2 J_1(kr_2) - r_1 J_1(kr_1)) + \\
  &(J_0(kr_2) - J_0(kr_1)) \cdot (r_2 Y_1(kr_2) - r_1 Y_1(kr_1)) ] \Big\},
\end{split}
\label{dissipation_factor_charge_Corbino}  
\end{equation}

\begin{equation}
  \mathrm{where} \qquad k(\sigma_{xx}/fC_\mathrm{TG}) \equiv \sqrt{\frac{2 \pi C_\mathrm{TG} f}{i A_\mathrm{TG} \sigma_{xx} }},
  \label{wave_number}
\end{equation}
and $J$ and $Y$ are Bessel functions of the first and second kind, respectively. (The detailed calculation is provided in Supplementary Information Section~VII).

To validate this model and extract the quantization, we performed two independent experiments, measuring the charge accumulation as a function of frequency ($f$) and then as a function of longitudinal conductance ($\sigma_{xx}$). The theoretical solution (Eq. \eqref{accumulated_charge_Corbino}) is fitted to these two datasets, with the intrinsic charge accumulation density (per unit AC field) $\eta_0^{(\mathrm{unit} B)} = \eta_0 / B_\mathrm{AC}$ serving as the primary fitting parameter, whose theoretical value is $e^2/h$. The geometric parameters of the Corbino disk ($r_1 = 350$ µm, $r_2 = 1000$ µm) are treated as known constants throughout the analysis.

We first measured and analyzed the frequency dependence of charge accumulation, with the results shown in Fig.~\ref{fig:fig3}\textbf{b}. To isolate frequency as the variable, we kept the longitudinal conductance stable around $\sigma_{xx} = 6.2 \times 10^{-7}$ S by adjusting the AC field amplitude at each frequency to maintain a stable sample temperature. Experimental data (scatter points with error bars displaying the standard error) were measured at DC fields of $\pm0.5$ T and $\pm1.0$ T. These fields were specifically chosen because they lie deeply within the QAH plateau, where $\sigma_{xx}$ is small and stable, ensuring reliable experimental conditions. The data from both fields are identical within experimental error, confirming the signal's stability. Before performing the fitting, we first compared the data to a theoretical prediction (solid lines). It was calculated using the expected quantized value $\eta_0^{(\mathrm{unit} B)} = e^2/h$ combined with independently measured parameters for $C_\mathrm{TG}$ (2.00 nF, see the Supplementary Information Section~VIII for details) and $\sigma_{xx}$ ($6.2 \times 10^{-7}$ S). Experimental data show agreement with the prediction within error, providing initial validation of our model. Furthermore, as highlighted by the inset, our measurements lie within the severe decay region, where the model correctly predicts the observed linear dependence of the quadrature component on frequency.

To quantitatively extract the intrinsic signal $\eta_0^{(\mathrm{unit} B)}$, we then fit the model (Eq. \eqref{accumulated_charge_Corbino}) to the data from both $\pm0.5$ T and $\pm1.0$ T, with the results summarized in Table~\ref{tab:fitting_results}. Our primary approach is a constrained, one-parameter fit (light solid lines), where the dissipation parameters $C_\mathrm{TG}$ and $\sigma_{xx}$ were fixed to their independently measured values, leaving only $\eta_0^{(\mathrm{unit} B)}$ as a free parameter. This fit was successful, yielding an $\eta_0^{(\mathrm{unit} B)}$ value consistent with the quantized value of $e^2/h$ with only a $5\%$ deviation. For comparison, we also performed a two-parameter fit (light dashed lines) where the dissipation term $\lambda_f = \frac{ \sigma_{xx} A_\mathrm{TG}}{C_\mathrm{TG}}$ was treated as a second free parameter. However, because the main data lie in a nearly linear regime, this less-constrained fit lacks the sensitivity to reliably separate the contributions from $\eta_0^{(\mathrm{unit} B)}$ and dissipation. This limitation highlights the importance of using accurately determined dissipation parameters. The success of the one-parameter fit, which relies on these independently measured values, therefore provides quantitative evidence that the intrinsic charge accumulation is indeed quantized at $e^2/h$.

As a second, independent validation of the model, we analyze the charge accumulation's dependence on longitudinal conductance, with the results shown in Fig.~\ref{fig:fig3}\textbf{c}. The experiment was performed at a fixed frequency of 277.777 Hz, while $\sigma_{xx}$ was tuned by varying the AC field amplitude, which controls the sample temperature via eddy current heating. $\sigma_{xy}$ was confirmed to remain quantized across the entire temperature range by simultaneously monitoring the Hall-bar sample. The charge accumulation signals, measured at DC fields of $\pm0.5$ T and $\pm1.0$ T, are plotted (scatter points) against $1/\sigma_{xx}$, which plays a role analogous to frequency in the decay model. The data are compared with a theoretical prediction (solid lines), which is calculated using the expected quantized value $\eta_0^{(\mathrm{unit} B)} = e^2/h$ and independently measured parameters ($C_\mathrm{TG} = 2.00$ nF, $f = 277.777$ Hz). The experimental data show agreement with this prediction, further validating the model. To quantitatively extract the intrinsic signal, we again apply the one-parameter fit to Eq. \eqref{accumulated_charge_Corbino}, leaving only $\eta_0^{(\mathrm{unit} B)}$ as the free parameter. The resulting fit (light solid lines), as detailed in Table~\ref{tab:fitting_results}, yields an intrinsic charge accumulation of $\eta_0^{(\mathrm{unit} B)} = 1.16 \pm 0.06$ $e^2/h$. This value is also in agreement with the theoretical quantum of $e^2/h$. The success of this one-parameter fit, performed on an independent data set, provides powerful, corroborating evidence that the underlying charge accumulation is indeed quantized.

\subsection*{Direct measurement of the quantized charge accumulation}

In the preceding measurements, the quantized QAH charge accumulation was obscured by strong dissipation and appeared primarily as a relatively small quadrature component. To directly access the quantized value of the charge accumulation, we further improved the measurement conditions by lowering the sample temperature to approximately 20 mK, thereby achieving an ultra-low sample longitudinal conductance of $\sigma_{xx} = 2.34\times 10^{-9}$ S (see Supplementary Information Section~IX for calibration details). In addition, we fabricated large-area simple disk-shaped samples (radius $r_0 = 1200$ µm) from the same batch as the Corbino disk sample to further reduce dissipation through geometric factors. For simple disk samples, Eqs. \eqref{accumulated_charge_Corbino} and \eqref{dissipation_factor_charge_Corbino}, which describe the charge signals, are replaced by
\begin{equation}
    \tilde{Q}_\mathrm{a.} / B_{\mathrm{AC}} =  g(\sigma_{xx}/fC_\mathrm{TG}) \, \pi r_0^2 \,\eta_0^{(\mathrm{unit} B)}
    \label{accumulated_charge_disk}
\end{equation}
and 
\begin{equation}
    g(\sigma_{xx}/f C_\mathrm{TG}) = 1 - \frac{2J_1(kr_0)}{kr_0J_0(kr_0)},
    \label{dissipation_factor_charge_disk}
\end{equation}
where the definition of $k(\sigma_{xx} /f C_\mathrm{TG})$ given in Eq. \eqref{wave_number} remains unchanged (see Supplementary Information of Ref. \cite{li2026} for details). The charge accumulation was still measured through the top gate, while the sample contact remained grounded.

\begin{figure}[h!]
  \centering
  \includegraphics[width=1\linewidth]{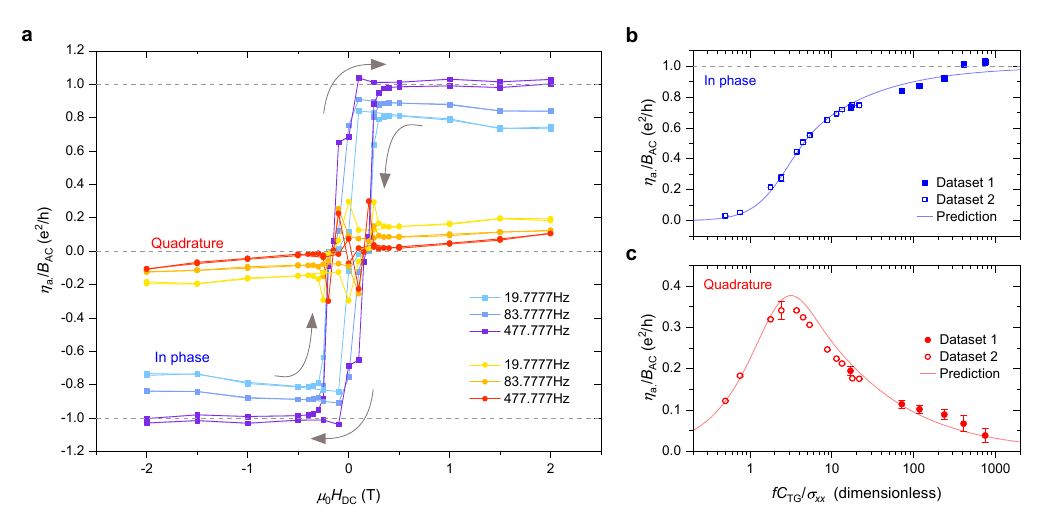}
  \caption{\textbf{Direct measurement of quantized charge accumulation and comprehensive parameter dependence of charge signals in simple disk samples.} \textbf{a}, DC-field dependence of $\eta_\mathrm{a.}/B_\mathrm{AC}$ from $f = 19.7777$ Hz to $f = 477.777$ Hz at an ultra-low $\sigma_{xx} = 2.34\times 10^{-9}$ S. In-phase and quadrature components are plotted separately. Arrows denote the direction of field sweeps. For clarity, only data at selected frequencies are shown.
  \textbf{b,c}, Dependence of the in-phase (\textbf{b}) and quadrature (\textbf{c}) components of $\eta_\mathrm{a.}/B_\mathrm{AC}$ on the combined parameter $f C_\mathrm{TG}/\sigma_{xx}$. The theoretical predictions are calculated using Eq.~\eqref{dissipation_factor_charge_disk}. Two experimental datasets are shown to cover a broad range of $f C_\mathrm{TG}/\sigma_{xx}$. The error bars on data points represent standard errors calculated from repeated measurements. Within each dataset, only $f$ is varied, while $\sigma_{xx}$ and $C_\mathrm{TG}$ are fixed. Datasets 1 and 2 were measured on disk samples I and II, respectively; the two samples have the same radius but different $\sigma_{xx}$ and $C_\mathrm{TG}$.}
  \label{fig:fig4}
\end{figure}

The DC-field dependence of the charge accumulation is shown in Fig.~\ref{fig:fig4}\textbf{a}, where data measured at three frequencies are presented. Within the QAH plateau, the in-phase component of $\eta_\mathrm{a.}/B_\mathrm{AC}$ at $f = 19.7777$ Hz is approximately $0.7\,e^2/h$, accompanied by a residual quadrature component of about $0.2\,e^2/h$. As the frequency increases, the in-phase component gradually approaches the quantized value, while the quadrature component is progressively suppressed toward zero. Eventually, at $f = 477.777$ Hz, we directly resolve an in-phase component that is quantized within experimental uncertainty, together with a quadrature component below $0.1\,e^2/h$.

After achieving nearly quantized charge accumulation at ultra-low $\sigma_{xx}$ and high $f$, we measured charge-accumulation signals within the QAH plateaus over a range of dissipation parameters to further validate our dissipation model (Fig.~\ref{fig:fig4}\textbf{b},\textbf{c}). Because the accessible frequency range of the AC magnetic field is limited, we used two $f$-dependent datasets with different $\sigma_{xx}$ and $C_\mathrm{TG}$ to cover a broad range of the combined parameter $f C_\mathrm{TG}/\sigma_{xx}$ and thereby test the full scaling behavior predicted by the model. The data are plotted as a function of $f C_\mathrm{TG}/\sigma_{xx}$, which combines the relevant independent parameters in the dissipation model (Eq.~\eqref{dissipation_factor_charge_disk}). $C_\mathrm{TG}$ and $\sigma_{xx}$ were independently calibrated from gate-to-contact impedance measurements (see Supplementary Information Section~IX for details). The detailed parameters of both datasets are summarized in Table~\ref{tab:fig4_parameters}.

Both experimental charge-accumulation datasets agree well with the theoretical predictions within experimental uncertainty. In the low $f C_\mathrm{TG}/\sigma_{xx}$ regime, the quadrature component dominates, consistent with the strongly dissipated charge accumulation measured in the Corbino disk geometry (Fig.~\ref{fig:fig3}\textbf{b},\textbf{c}). As $f C_\mathrm{TG}/\sigma_{xx}$ increases, the in-phase component increases monotonically, whereas the quadrature component exhibits a single-peaked dependence. At sufficiently large $f C_\mathrm{TG}/\sigma_{xx}$, the effect of dissipation on charge accumulation is suppressed to a negligible level and the signal approaches the ideal response. The in-phase component saturates toward $e^2/h$, reaching the quantized value within $\pm 5\%$, whereas the quadrature component decreases toward zero in the same limit.

In summary, by further suppressing $\sigma_{xx}$ and modifying the sample geometry, we substantially reduce the dissipation of charge accumulation and achieve a direct and reliable measurement of the quantized charge accumulation signal. By covering a broad range of the combined dissipation parameter $f C_\mathrm{TG}/\sigma_{xx}$, we comprehensively validate the dissipation mechanism described by our model (Eq.~\eqref{dissipation_factor_charge_disk}). 

\section*{Discussion}

In conclusion, we have experimentally validated an out-of-plane capacitive method that enables quantitative detection of the field-induced quantized charge accumulation in a topological QAH system. This validation holds across both high and low dissipation regimes, with experimental data closely matching theoretical predictions. Notably, the achieved charge resolution is ultrahigh, reaching below $0.1\ \mathrm{fC/Gs}$, underscoring the sensitivity of this technique.

The critical advantage of this out-of-plane method lies in its ability to access the defining bulk property of 3D TIs: the topological magnetoelectric effect (TME). The TME is a condensed-matter manifestation of the 4D QHE \cite{SCZhang2001_4DQH, qi2008topological, Vanderbilt2009_3DTI}, characterized by the relationship $\Delta P = \frac{e^2}{2h}N_\mathrm{Ch}^{(2)}\Delta B$, where $N_\mathrm{Ch}^{(2)}$ is the second Chern number and the four-dimensional parameter space combines three spatial dimensions with time. This effect emerges in a 3D TI when its surface magnetizations are uniformly oriented either all inward or all outward, creating an axion insulator (AI) state \cite{Mogi2017_Axion_modulation_CBST, Mogi2017_Axion_CVBST, ChangCZ2018_Axion_CVBST, WangYayu2020_Axion_MBT, XuYang2014_Axion_BSTS, ChangCZ2023_Axion_CVBST100nm, FengXiao2024_Axion_MBE_MBT}. In a thin AI film subjected to a perpendicular magnetic field, this polarization difference manifests as a quantized surface charge accumulation of opposite signs: $+\frac{e^2}{2h}\Delta B$ on one surface and $-\frac{e^2}{2h}\Delta B$ on the other.

Critically, this signature is fundamentally inaccessible to conventional in-plane measurement techniques, including transport \cite{1992_Russia_GaAs_chargepumping&Streda, RIKEN_CBST_chargepumping} and optical methods \cite{WuLiang2016_3DTI_THz_optical, Tokura2016_QAH_THz_optical, Shimano2010_QHE_THz_optical, Shimano2013_graphene_THz_optical}. When applied to an AI, these techniques inherently detect the sum of the top and bottom surface responses, which strictly cancels to zero (see Table~S1 in the Supplementary Information). Therefore, accessing the 4D QHE necessitates the out-of-plane capacitive approach demonstrated here. In a dual-gate device, the TME surface charge of opposite signs would be detected as an out-of-plane current, $I_{\mathrm{TME}} \propto \gamma_\mathrm{geo} \frac{e^2}{2h}N_\mathrm{Ch}^{(2)}\Delta B$, where $\gamma_\mathrm{geo}$ is a geometric attenuation factor.

It is important to distinguish the physics observed here from the target TME. Our observation of field-induced quantized charge accumulation in the QAH state \cite{ChangCZ2013_QAH_CBST, KouXufeng2014_QAH_CBST, Tokura2014_QAH_CBST, ZhangYuanbo2020_QAH_MBT} of a 3D TI stems from 2D QHE physics of the surface state, not the 4D QHE physics of the bulk. In the QAH state of a 3D TI, the coexistence of multi-domain states with $N_\mathrm{Ch}^{(2)} = \pm 1$ cancels the total bulk electric polarization difference, resulting in surface charge accumulation of the same sign on both surfaces. In contrast, the AI state---being a single-domain 4D QHE state---manifests in charge accumulation of opposite signs on the two surfaces. While both states exhibit quantized charge accumulation on a single surface, their underlying mechanisms and experimental signatures differ fundamentally.

Nevertheless, the validation of our methodology has direct implications for the detection of the 4D QHE. We have demonstrated sensitive surface charge measurement and quantitative modeling to overcome inherent dissipation effects. While our single-gate geometry detects the net charge accumulation of the QAH state (a 2D QH system) with a current signal $I_{\mathrm{QAH}} \propto \pm\frac{e^2}{h} \Delta B$, a dual-gate structure for an AI state would detect the TME's quantized electric polarization difference, a signature of the 4D QHE, with a current signal $I_{\mathrm{TME}} \propto \gamma_\mathrm{geo} \frac{e^2}{2h}N_\mathrm{Ch}^{(2)}\Delta B$. The geometric attenuation factor $\gamma_\mathrm{geo} \ll 1$ presents a significant experimental challenge; however, our work, when combined with active capacitive compensation to enhance $\gamma_\mathrm{geo}$ \cite{li2026} and ultrasensitive capacitance calibration to determine $\gamma_\mathrm{geo}$, offers a promising pathway toward directly detecting the TME and exploring the 4D QHE in condensed matter physics.

\section*{Methods}

\subsection*{Thin-film growth and device fabrication}

The devices, including the Corbino disk, simple disk, and Hall bar geometries, were fabricated from a 6-QL CBST film grown by MBE on a GaAs substrate. To minimize sample degradation, all patterning steps were performed using Molybdenum (Mo) shadow masks in a solvent-free process. First, the film was patterned via argon ion milling to define the device mesa. Subsequently, Ti/Au (3 nm/47 nm) contact electrodes were deposited by thermal evaporation. The inner and outer radii of the Corbino disk sample's contact electrodes are 350 µm and 1000 µm, respectively, corresponding to a sample area of 2.757 $\mathrm{mm^2}$. The radii of both simple disk samples are 1200 µm. A thick AlO$_x$ gate dielectric (about 140 nm for the Corbino disk sample, 200 nm for the simple disk sample I, and 270 nm for the simple disk sample II) was then deposited at room temperature using oxygen plasma-assisted atomic layer deposition (ALD). Finally, Ti/Au (8 nm/72 nm) top gates were evaporated. For the Corbino disk sample, the top gate partially overlaps the area of the contact electrodes (90 µm width) to fully cover the entire sample, which has a total area of 3.520 $\mathrm{mm^2}$. The insulation of the large-area top gate was ensured by the thick AlO$_x$ dielectric and a smooth topography at the edges of contact electrodes. This smooth topography, which mitigates defects in the overlying dielectric, is the result of the shadow mask method coupled with sample rotation during metal evaporation.

\subsection*{In-plane transport measurement}

The Hall resistance ($\rho_{yx}$) and longitudinal resistance ($\rho_{xx}$) of the Hall-bar sample were measured using a standard low-frequency AC lock-in technique. An excitation current of 10 nA (13.777 Hz) was applied to the sample, calibrated via a 100 $\mathrm{k\Omega}$ series resistor. The resulting transverse and longitudinal voltages were detected using lock-in amplifiers (Stanford Research Systems, models SR830 and SR865A). The Hall conductance ($\sigma_{xy}$) and longitudinal conductance ($\sigma_{xx}$) were then obtained by inverting the resistivity tensor. 

For the Corbino disk sample, the longitudinal conductance was determined using a two-terminal, constant-voltage method. A 5 µV AC (13.777 Hz) voltage ($U_c$) was applied between the inner and outer contacts, and the resulting current ($I_c$) was measured with a custom-built transimpedance current preamplifier ($10^7$ V/A gain, based on an ADA4530-1 chip) connected to a lock-in amplifier (model SR830). The conductance was calculated using the standard Corbino formula: $\sigma_{xx} = \frac{\ln(r_2/r_1)}{2\pi} I_c/U_c$, where $r_1$ and $r_2$ are the radii of the inner and outer contacts, respectively.

\subsection*{AC magnetic field generation}

The AC magnetic field was generated by two identical small coils arranged to clamp the sample holder. The entire coil assembly was anchored to the cold finger of the 1 K still plate in the dilution refrigerator, ensuring that the AC and DC fields were coaxial and perpendicular to the sample plane. Each coil was constructed by winding a 0.438 mm diameter Cu-clad NbTi superconducting wire onto a brass bobbin. The solenoid has a total of approximately 340 turns, with an inner diameter of 24 mm and a length of 15 mm. The two coils were mounted coaxially with a separation of 11 mm to form a Helmholtz pair, ensuring high field homogeneity near the sample. To shield against capacitive coupling, each coil was wrapped with a 0.2 mm thick copper foil (as a Faraday cage). The two coils were wired in series, and this pair was driven by the differential sine outputs of an SR865A lock-in amplifier (50 $\Omega$ output impedance per terminal), providing a zero-mean-voltage excitation to minimize measurement crosstalk.

The amplitude and phase of the resulting AC magnetic field were calibrated in-situ at each measurement frequency in a static DC field of 0.01 T. The calibration was performed using a Hall sensor fabricated from a GaAs/AlGaAs heterostructure containing a 2D electron gas. For the calibration, a 1 µA DC current was applied to pass through the sensor, and the resulting AC Hall voltage was measured with a lock-in amplifier. The AC field was then determined by comparing this AC voltage to the sensor's calibrated DC Hall coefficient ($k_\mathrm{Hall} = \Delta \rho_{yx}/\Delta B$). To remove background crosstalk in the AC voltage, this process was repeated with a reversed DC current (-1 µA) and the results were antisymmetrized. The complete, calibrated AC field response per volt of excitation is presented in the Supplementary Information Fig.~S10. In order to maintain a low and constant sample temperature, we used small and frequency-dependent AC field amplitudes, as shown in the Supplementary Information Fig.~S5. The range of AC field amplitude in the experiment was 0.033 Gs to 0.180 Gs (637.777 Hz to 83.777 Hz, root-mean-square (RMS) value).

\subsection*{Charge signal measurement}

The charge accumulation induced by the AC magnetic field was measured on the Corbino disk sample (Fig.~\ref{fig:fig1}\textbf{b,e}) and the simple disk sample. The resulting AC signal was detected as a current from the top gate, which was connected to the virtual ground input of a custom-built current preamplifier ($10^7$ V/A gain, based on an ADA4530-1). This configuration ensures the measured current corresponds directly to the charge variation within the sample. The amplifier's output was measured by a lock-in amplifier, referenced to the AC field's source (an SR865A). The preamplifier was located at room temperature and connected to the sample's top gate via a low-resistance coaxial cable. Because the amplifier's virtual ground holds the cable's core conductor at the same potential as its grounded shield, the parasitic capacitance of the cable (about 900 pF, same magnitude as the $C_\mathrm{TG}$ of the sample) has a negligible effect on the measurement.

For the Corbino disk sample, a DC gate voltage was applied to tune the sample's Fermi level (Fig.~\ref{fig:fig1}\textbf{e}), while the simple disk sample has a well tuned Fermi level with no need for gate voltage. The optimal gate potential of +9.4 V (gate relative to sample) was identified as the point of minimum longitudinal conductance. To isolate the DC gating circuit from the AC measurement, a bias voltage was applied to the sample's contacts while the gate remained at ground potential. Specifically, both Corbino contacts were biased to -9.4 V using a National Instruments PXIe-4322 module, with a 100 k$\Omega$ series resistor for current protection. To provide a low-impedance path for AC currents, a 10 µF bypass capacitor was connected between the sample contacts and ground. This capacitor effectively shorts the gating circuit at the measurement frequencies (e.g., presenting impedance of only 190 $\Omega$ at 83.777 Hz) while maintaining the desired DC potential on the sample.

Since the charge accumulation is sensitive to the longitudinal conductance and corresponding temperature, we measured signals at several DC field points rather than continuously sweeping the field, which would generate extra temperature variation. Each point was held for 900 seconds to ensure a stable temperature, and the data from the last 300 seconds were recorded.

\backmatter

\section*{Data Availability}
The data generated in Figs. 1-4 of this study have been provided in the source data file. Source data are provided with this paper. All other data that support the plots within this paper are available from the corresponding author upon reasonable request.

\bigskip

\bibliography{MainPaper}

\section*{Acknowledgements}
T.L. acknowledges the support for the project by the National Key R\&D Program of China (No. 2021YFA1401600). Y.Liu acknowledges the support by the National Key R\&D Program of China (Grant No. 2021YFA1401900). We thank Xing He and Wanjun Jiang for providing access to their Ar ion etching system, which was used for sample fabrication. We also thank Daiqiang Huang for his discussions with us on the cryogenic coil structure. Y.Li thanks Lewei Li for her help with the schematic preparation.

\section*{Author contributions}
T.L. conceived and designed the project, provided overall direction, supervised the experiments, and coordinated collaborations among the research groups. Y.Li and J.C. designed the ultrahigh-sensitivity experimental setup for charge accumulation detection and performed the measurements with extensive assistance from R.W. and Y.Liu, as well as through in-depth discussions with T.L. R.W. and Y.Liu developed the custom ultrahigh-sensitivity current preamplifier. The high-quality QAH thin films were grown by Y.Z. and X.K. and subsequently characterized and fabricated by Y.Li with support from J.C., Y.D., and J.X. Analysis of the data was conducted by Y.Li and T.L. The manuscript was written by Y.Li and T.L. with input from all authors. All authors discussed the results and provided feedback for the manuscript.

\section*{Competing interests}
The authors declare no competing interests.

\section*{Tables}

\begin{table}[h!]
\centering
\caption{Fitting results for the intrinsic charge accumulation. The uncertainty represents the 95\% confidence interval.}
\label{tab:fitting_results}
\begin{tabular}{l@{\hspace{1.5em}}ccc} 
\toprule
Experiment & Parameter & Fitted Value & Theoretical Value \\ 
\midrule
\multicolumn{4}{l}{A. Frequency Dependence} \\
\quad 1-Parameter Fit$^a$     & $\eta_0^{(\mathrm{unit} B)}$ ($e^2/h$)      & $1.05 \pm 0.06$       & 1                        \\
\cmidrule(lr){2-4}
\quad \multirow{2}{*}{2-Parameter Fit} & $\eta_0^{(\mathrm{unit} B)}$ ($e^2/h$)      & $0.69 \pm 0.20$       & 1                        \\
\quad                     & $\lambda_f$ (10$^{-3}$ m$^2$/s) & $0.71 \pm 0.20$       & 1.09$^b$                 \\ 
\midrule
\multicolumn{4}{l}{B. $\sigma_{xx}$ Dependence} \\
\quad 1-Parameter Fit$^c$     & $\eta_0^{(\mathrm{unit} B)}$ ($e^2/h$)      & $1.16 \pm 0.06$       & 1                        \\ 
\bottomrule
\multicolumn{4}{l}{\footnotesize{$^a$Independently measured values are used: $\sigma_{xx}=6.2\times 10^{-7}$ S, $A_\mathrm{TG}=3.52\times 10^{-6}$ m$^2$, and $C_\mathrm{TG}=2.00$ nF.}} \\
\multicolumn{4}{l}{\footnotesize{$^b$Theoretical value of $\lambda_f = \sigma_{xx}A_\mathrm{TG}/C_\mathrm{TG}$ is calculated from independently measured values represented above$^a$.}}\\
\multicolumn{4}{l}{\footnotesize{$^c$Independently measured values are used: $f = 277.777$ Hz, $A_\mathrm{TG}=3.52\times 10^{-6}$ m$^2$, and $C_\mathrm{TG}=2.00$ nF.}}
\end{tabular}
\end{table}

\begin{table}[h!]
\centering
\caption{
Parameters for the two $f$-dependent datasets of charge accumulation.
}
\label{tab:fig4_parameters}
\begin{tabular}{cccccc}
\toprule
Dataset & Sample & $\mu_0 H_\mathrm{DC}$ (T) & $C_\mathrm{TG}$ (nF) & $\sigma_{xx}$ (S) & $f$ range (Hz) \\
\midrule
1 & Disk sample I  & 1.5 & 2.00 & $2.34\times10^{-9}$ & 19.7777--877.777 \\
2 & Disk sample II & 0.5 & 1.51 & $2.35\times10^{-8}$ & 7.7977--337.777 \\
\bottomrule
\end{tabular}
\end{table}

\section*{Figure Legends}

\end{document}


\title{Supplementary Information for ``Observation of Quantized Charge Accumulation in a Quantum Anomalous Hall System''}

\maketitle
\tableofcontents
\clearpage


\section{Optimal gate voltage of Hall bar and Corbino disk samples}
\label{sec:SMsection1}

\begin{figure}[htbp]
  \centering
  \includegraphics[width=1\linewidth]{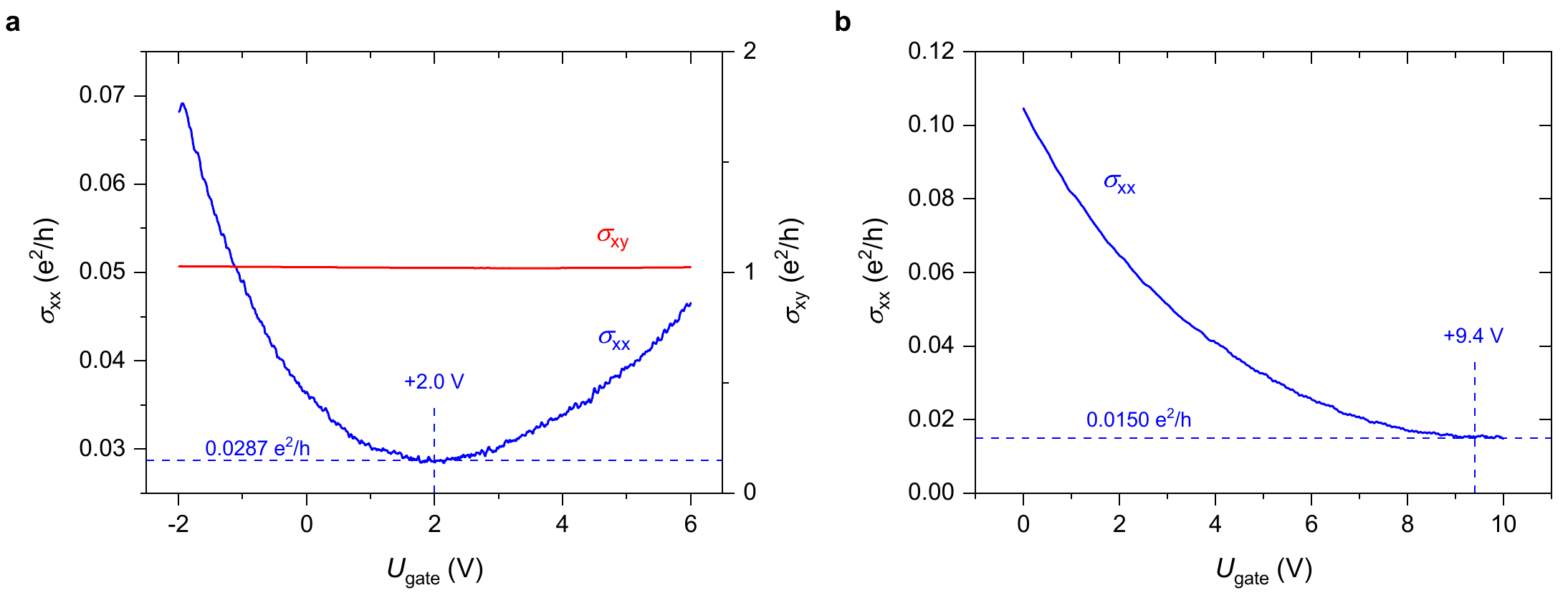}
  \caption{\textbf{The gate voltage dependence of conductances for Hall bar and Corbino disk samples.} \textbf{a}, The gate voltage dependence of $\sigma_{xx}$ and $\sigma_{xy}$ for Hall-bar sample at 1.5 T field. \textbf{b}, The gate voltage dependence of $\sigma_{xx}$ for Corbino disk sample at 0.8 T field.}
  \label{fig:figS1}
\end{figure}

The optimal voltage of the QAH sample is determined by the minimum longitudinal conductance, while the Hall conductance of the sample maintains quantization within a wide range of gate voltage. The longitudinal conductance $\sigma_{xx}$ of the Hall-bar sample was measured by conventional voltage methods with constant current \cite{ChangCZ2013_QAH_CBST, KouXufeng2014_QAH_CBST, Tokura2014_QAH_CBST}, while the $\sigma_{xx}$ of the Corbino disk sample was calculated from two-terminal resistance \cite{RIKEN_CBST_chargepumping}, where a constant voltage was applied between the inner and outer contacts and we measured the current passing through. The gate voltage dependence of conductances for the two types of samples is represented in Fig.~\ref{fig:figS1}\textbf{a} and Fig.~\ref{fig:figS1}\textbf{b}. The measurements were conducted at DC fields of 1.5 T and 0.8 T respectively, both in the field region of full quantization for the QAH effect. (The difference in field is due to other data comparison requirements and does not affect the value of the optimal gate voltage). The measurement temperature was about 200 mK. The optimal gate voltage of the Hall-bar sample is +2.0 V with a minimum $\sigma_{xx}$ of 0.0287 $e^2/h$, while the optimal gate voltage of the Corbino disk sample is +9.4 V with a minimum $\sigma_{xx}$ of 0.0150 $e^2/h$. This difference comes from the inhomogeneity of the sample film. In charge accumulation and charge pumping measurements, this optimal gate voltage was always applied, despite the different approaches.

\section{Antisymmetrization of charge accumulation signals against the DC magnetic field}
\label{sec:SMsection2}


\begin{figure}[htbp]
  \centering
  \includegraphics[width=1\linewidth]{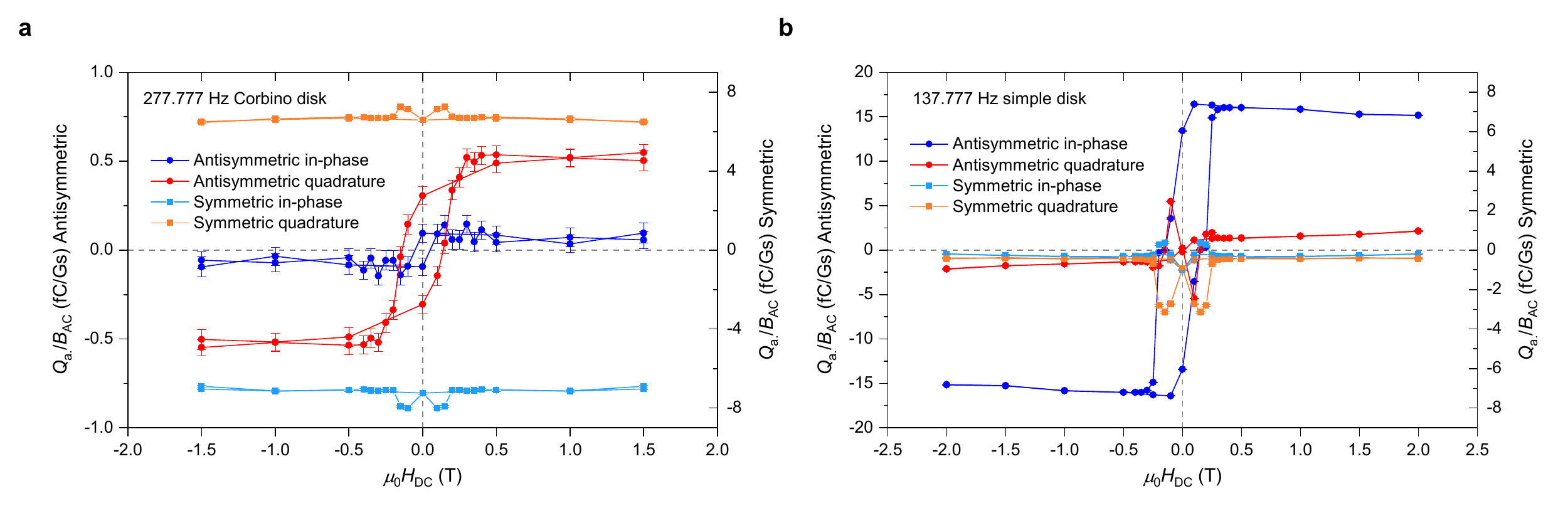}
  \caption{\textbf{Symmetric and antisymmetric components of charge accumulation signals.} \textbf{a}, Data measured at 277.777 Hz on the Corbino disk sample (strong dissipation regime). \textbf{b}, Data measured at 137.777 Hz on the simple disk sample I (weak dissipation regime). The error bars on data points represent standard errors calculated from repeated measurements.}
  \label{fig:figS2}
\end{figure}

The DC field dependence of the symmetric and antisymmetric components of charge accumulation signals is shown in Fig.~\ref{fig:figS2}. In-phase and quadrature components are plotted separately. Data measured on the Corbino disk sample at $f = 277.777$ Hz, $B_\mathrm{AC} = 0.062$ Gs, and on the simple disk sample I at $f = 137.777$ Hz, $B_\mathrm{AC} = 0.099$ Gs are plotted as examples. The antisymmetrized signals $Q^\mathrm{antisymm}$ are calculated by $Q^\mathrm{antisymm}_{ \uparrow \mathrm{or} \downarrow}(B) = \left[Q_{ \uparrow \mathrm{or} \downarrow}(B) - Q_{ \downarrow \mathrm{or} \uparrow}(-B)\right]/2$, and symmetrized signals $Q^\mathrm{symm}$ are calculated by $Q^\mathrm{symm}_{ \uparrow \mathrm{or} \downarrow}(B) = \left[Q_{ \uparrow \mathrm{or} \downarrow}(B) + Q_{ \downarrow \mathrm{or} \uparrow}(-B)\right]/2$, where $B$ represents the value of the DC magnetic field and $\uparrow$ or $\downarrow$ represents the sweep direction of the field.

This procedure effectively isolates the genuine physical signal from experimental artifacts. The antisymmetric component corresponds to the true charge accumulation signal, while the symmetric component represents the instrumental crosstalk. This crosstalk is insensitive to the field, except for field points where $\sigma_{xx}$ obviously increases. The rationale for this separation is grounded in the physical origin of the crosstalk. It arises from a parasitic Faraday voltage (theoretically independent of the DC field) on the measurement wires, which induces a spurious charge signal via the measurement impedance related to the sample's longitudinal conductance, $\sigma_{xx}$. Since $\sigma_{xx}$ is symmetric with the DC field (i.e., $\sigma_{xx}^{ \uparrow \mathrm{or} \downarrow}(B) = \sigma_{xx}^{ \downarrow \mathrm{or} \uparrow}(-B)$), the resulting crosstalk is also symmetric. This intrinsic symmetry of the artifact allows it to be reliably separated from the genuine, antisymmetric charge accumulation signal. Furthermore, because dissipation is well suppressed in the disk sample, the ratio of crosstalk background to signal is reduced by two orders of magnitude compared to the Corbino sample.

\section{Measurement of Laughlin charge pumping in the Corbino disk sample}
\label{sec:SMsection3}


\begin{figure}[htbp]
  \centering
  \includegraphics[width=1\linewidth]{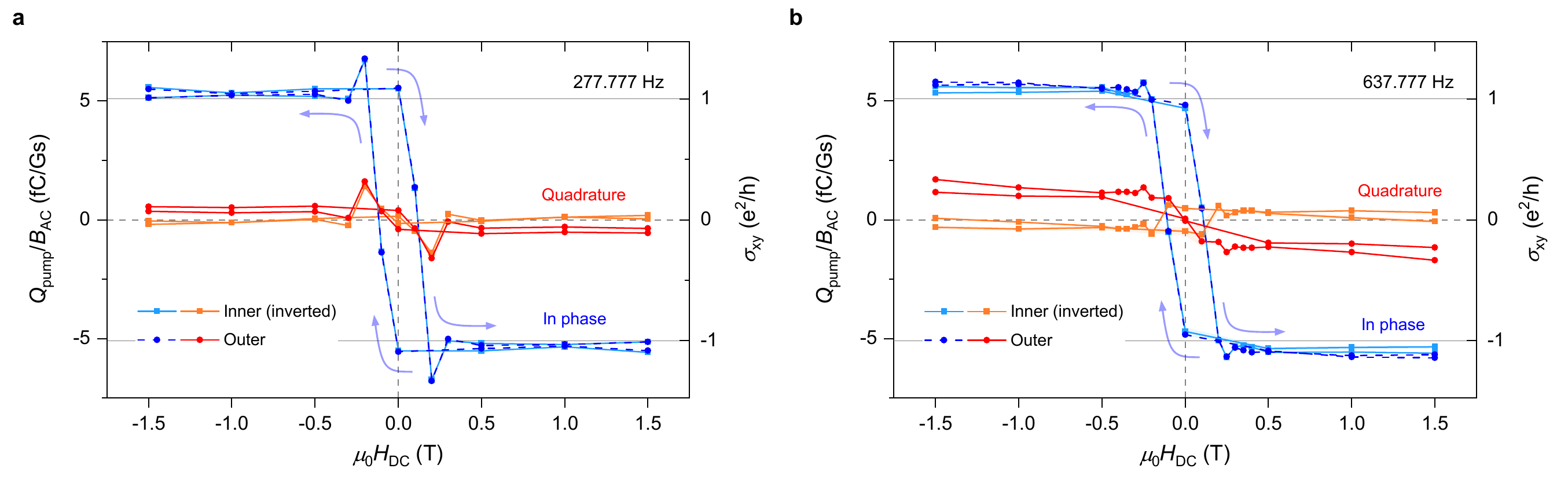}
  \caption{\textbf{The Laughlin charge pumping of the Corbino disk sample at inner and outer contacts.} The right vertical axis shows the corresponding Hall conductance $\sigma_{xy}$ calculated from the in-phase charge signal. The arrows show the sweep direction of the DC field related to the in-phase data. The inner contact charge signals have been inverted for a convenient comparison. \textbf{a}, Data measured at a frequency of 277.777 Hz. (Same data as in Fig. 2\textbf{b} of the main text.) \textbf{b}, Data measured at a frequency of 637.777 Hz.} 
  \label{fig:figS3}
\end{figure}

The Laughlin charge pumping in the Corbino disk sample was measured using a current-based method analogous to the charge accumulation experiments. Both inner and outer contact electrodes were connected to current preamplifiers to directly detect the pumped charge. The top gate was AC-grounded at the optimal DC voltage for gating (identical circuit to the charge accumulation setup), ensuring consistent experimental conditions with the measurement of charge accumulation. Charge pumping signals in response to AC fields at two frequencies (277.777 Hz and 637.777 Hz) are shown in Fig.~\ref{fig:figS3}\textbf{a} and Fig.~\ref{fig:figS3}\textbf{b}, with the corresponding Hall conductance $\sigma_{xy}$ plotted on the right vertical axes. The Hall conductance is calculated as $\sigma_{xy} = \frac{2 \ln{(r_2/r_1)}}{\pi (r_2^2-r_1^2)} \frac{Q_\mathrm{pump}^\mathrm{in-phase}}{B_\mathrm{AC}}$ \cite{RIKEN_CBST_chargepumping}, which assumes fully dissipated conditions (no charge accumulation) for a Corbino disk with inner radius $r_1 = 350$ µm and outer radius $r_2 = 1000$ µm. This formula remains applicable here, as charge accumulation at these frequencies is small and primarily in quadrature with the pumping signal, introducing negligible deviation. Anomalous signals at fields of 0.15 T (upward) and -0.15 T (downward)—within the QAH sample's magnetization flipping range (0.1 T to 0.2 T)—were removed due to dominant antisymmetric crosstalk arising from the large and unstable $\sigma_{xx}$.

The in-phase components of charge pumping exhibit perfect quantization on the QAH plateaus, including at zero field, directly manifesting the 2D Hall conductance. Differences in pumping signals between inner and outer contacts are evident, particularly in the quadrature component. This difference corresponds to the investigated charge accumulation within the sample. A detailed comparison between this charge difference and the charge accumulation directly detected from the top gate is provided in Supplementary Information (SI) Section~\ref{sec:SMsection4}.

Our charge pumping results align with a strongly dissipative regime, where inner and outer contacts pump nearly equal charge. We can further theoretically analyze the undissipated Laughlin charge pumping in the Corbino disk sample, where the accumulated charge stays on the sample surface and different amounts of charge are pumped at two contacts. In this ideal case, the pumped charge is strictly proportional to the magnetic flux change enclosed by the edge, as $Q^\mathrm{inner}_\mathrm{pump} = - \pi r_1^2 \Delta B \sigma_{xy}$ and $Q^\mathrm{outer}_\mathrm{pump} = \pi r_2^2 \Delta B \sigma_{xy}$ (defining the sign of inward charge as positive), while the value of the charge accumulation is $Q_\mathrm{a.} = \pi(r_2^2-r_1^2) \Delta B \sigma_{xy}$. The whole system obeys charge conservation $Q^\mathrm{inner}_\mathrm{pump} + Q^\mathrm{outer}_\mathrm{pump} = Q_\mathrm{a.}$. The charge dissipation can be modeled as charge diffusion from the sample to contacts, equalizing pumped charge under strong dissipation. Quantitative analysis of finite-dissipation cases is detailed in SI Section~\ref{sec:SMsection7}.

\section{Difference in Laughlin charge pumping signals between inner and outer contacts (compared with direct measurement of charge accumulation)}
\label{sec:SMsection4}

\begin{figure}[htbp]
  \centering
  \includegraphics[width=1\linewidth]{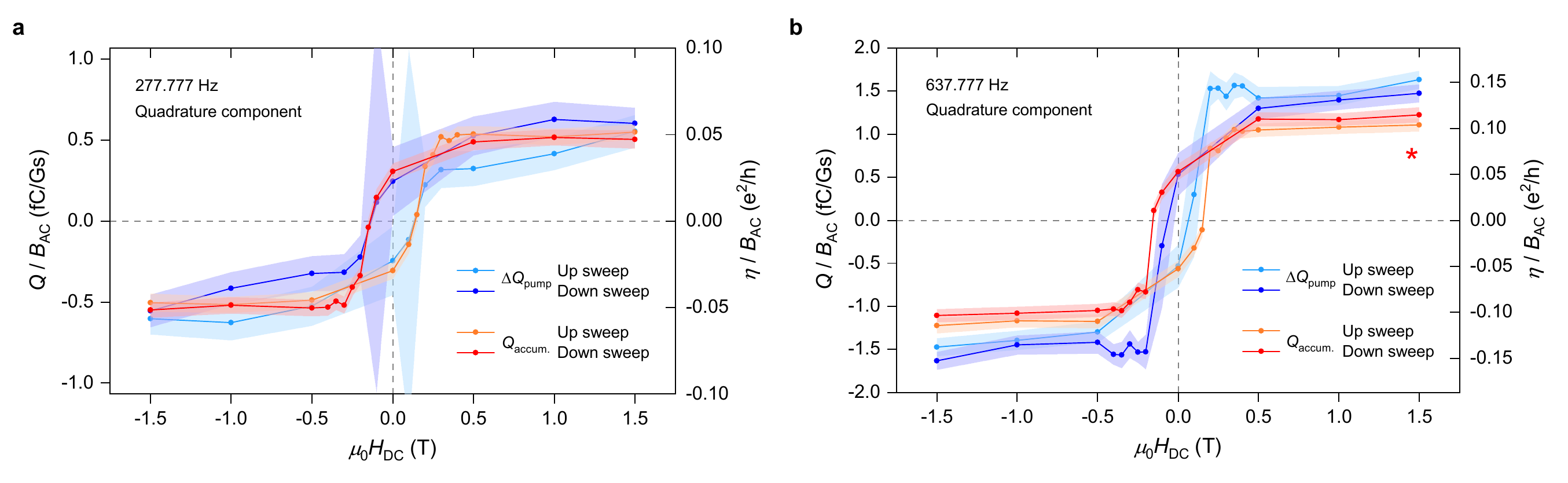}
  \caption{\textbf{Comparison of the directly measured charge accumulation ($Q_\mathrm{accum.}$) and the difference in charge pumping signals between the two contacts of the Corbino disk sample ($\Delta Q_\mathrm{pump}$).} Both datasets are plotted as a function of the DC magnetic field, with shaded areas representing the standard error of the data points. Data from different field sweep directions are shown in different colors. *The charge accumulation signal at 637.777 Hz is slightly smaller than the value under expected conditions, likely due to temperature fluctuations. (The real value remeasured on QAH plateaus is approximately 1.4 fC/Gs, as shown in main text Fig. 3\textbf{b}, which aligns better with $\Delta Q_\mathrm{pump}$.) \textbf{a}, Data at 277.777 Hz. \textbf{b}, Data at 637.777 Hz.}
  \label{fig:figS4}
\end{figure}

The difference in charge pumping signals between the inner and outer contacts of the Corbino disk sample corresponds to the charge accumulation within the sample. Comparisons between this charge difference ($\Delta Q_\mathrm{pump}$) and the charge accumulation directly measured from the top gate ($Q_\mathrm{accum.}$) are shown in Fig.~\ref{fig:figS4}\textbf{a} and Fig.~\ref{fig:figS4}\textbf{b}. Data measured at both 277.777 Hz and 637.777 Hz demonstrate consistent results between $\Delta Q_\mathrm{pump}$ and $Q_\mathrm{accum.}$ within the experimental error, validating the charge accumulation model. Notably, the error in the direct charge measurement is smaller than that in the charge pumping difference, especially when the longitudinal conductance $\sigma_{xx}$ is large. This ensures that the direct charge measurement provides better resolution (e.g., clearer hysteresis loops at low magnetic fields), highlighting the technical advantage of our out-of-plane method.

\section{Frequency dependence of AC field amplitudes for a constant sample temperature}
\label{sec:SMsection5}


\begin{figure}[htbp]
  \centering
  \includegraphics[width=0.8\linewidth]{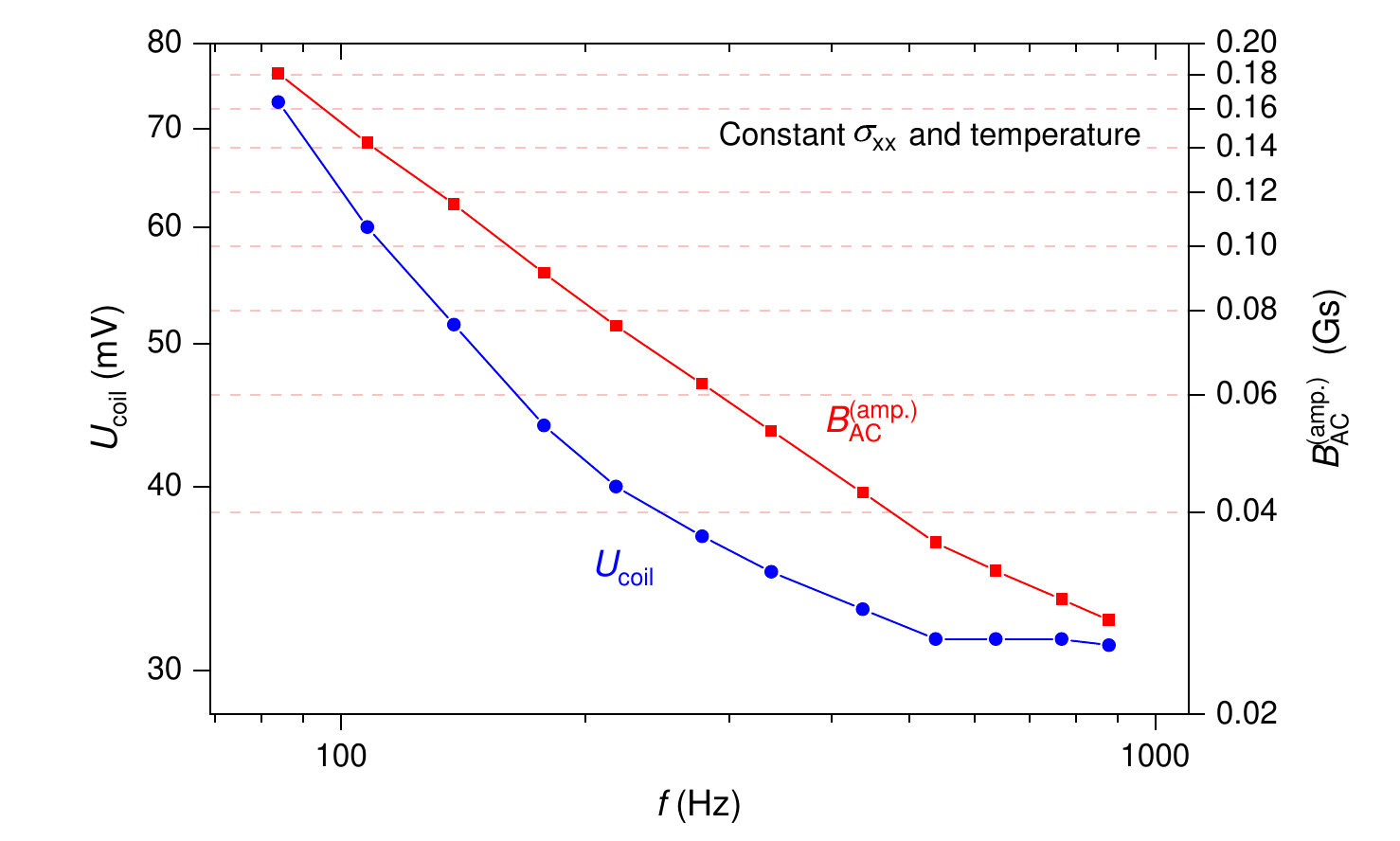}
  \caption{\textbf{Frequency dependence of source voltage amplitude $U_\mathrm{coil}$ and corresponding AC field amplitude $B_\mathrm{AC}^\mathrm{amp.}$ used to maintain a constant sample temperature.} Logarithmic coordinates are used on all axes. The longitudinal conductance $\sigma_{xx}$ of the Corbino disk sample was held within $(6.23 \pm 0.02) \times 10^{-7}$ S at a DC field of 0.8 T.}
  \label{fig:figS5}
\end{figure}

Eddy current heating from the AC magnetic field strongly influences the temperature, which alters $\sigma_{xx}$ and the charge dissipation of the QAH sample. To isolate frequency effects from the change of $\sigma_{xx}$, we adjusted the AC field amplitude at each frequency to ensure consistent heating and thus a fixed sample temperature. This was achieved by monitoring the two-terminal resistance of the Corbino disk sample at a DC field of 0.8 T and optimal gate voltage (+9.4 V). The resistance was maintained at $(268 \pm 1)$ $\mathrm{k \Omega}$, corresponding to a longitudinal conductance $\sigma_{xx}$ of $(6.23 \pm 0.02) \times 10^{-7}$ S, via adjustments to the source voltage amplitude $U_\mathrm{coil}$ of the AC field. The resulting $U_\mathrm{coil}$ and corresponding AC field amplitude $B_\mathrm{AC}^\mathrm{amp.}$ values are plotted in Fig.~\ref{fig:figS5} as a function of frequency. The figure shows a clear trend where the source voltage amplitude and AC field amplitude decrease with increasing frequency, reflecting the compensation for frequency-dependent heating to maintain constant $\sigma_{xx}$. 

\section{Out-of-plane charge measurement scheme for QAH total charge accumulation and TME polarization}
\label{sec:SMsection6}

\begin{figure}[htbp]
  \centering
  \includegraphics[width=0.8\linewidth]{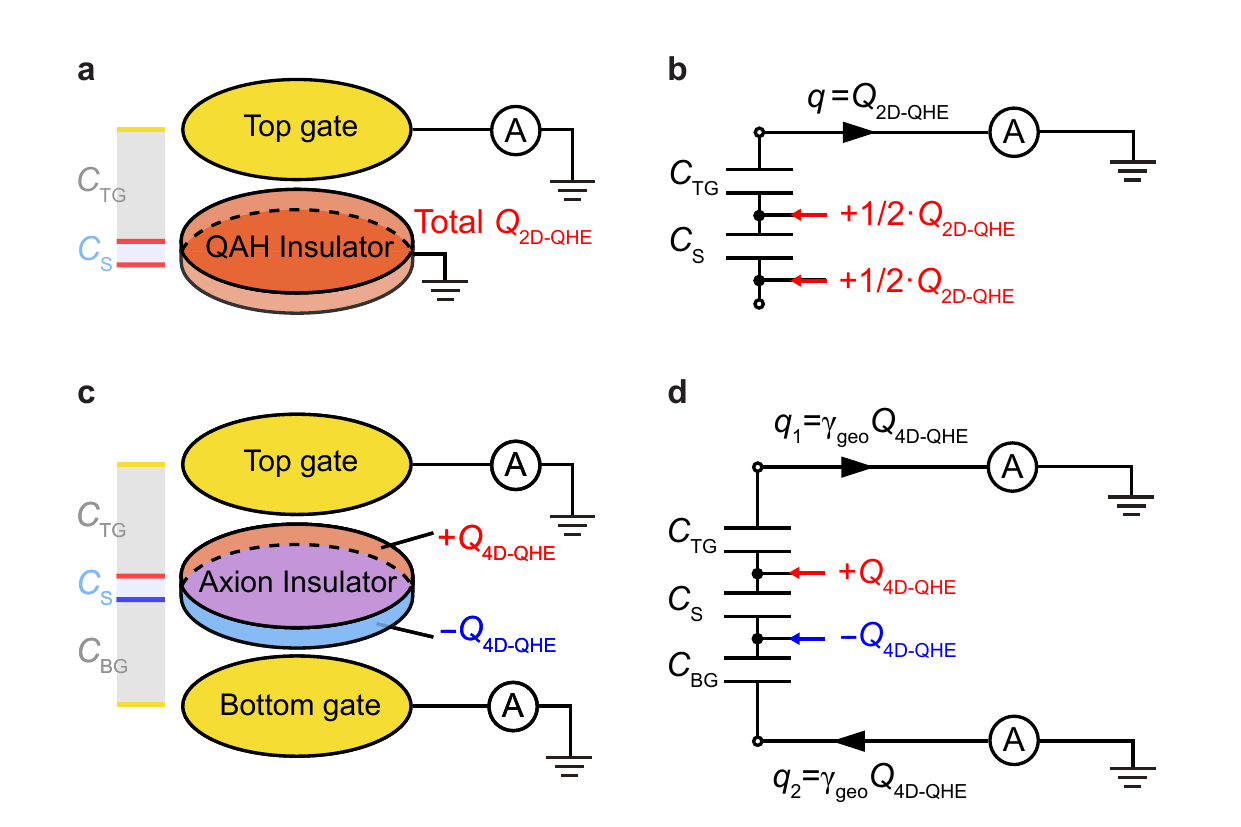}
  \caption{\textbf{Out-of-plane charge measurement schemes for QAH total charge accumulation and TME polarization.}
  \textbf{a,b}, Schematic and effective circuit for measuring the field-induced total charge accumulation in the QAH state through a single top gate. The intrinsic total accumulated charge is $Q_\mathrm{2D\mathrm{-}QHE}=A_\mathrm{S}\Delta B \cdot \frac{e^2}{h}$, and the measured charge signal is directly $Q_\mathrm{2D\mathrm{-}QHE}$.
  \textbf{c,d}, Schematic and effective circuit for measuring the TME polarization through dual gates. The intrinsic polarization produces opposite surface charges $\pm Q_\mathrm{4D\mathrm{-}QHE}=\pm A_\mathrm{S}\Delta B \cdot \frac{e^2}{2h}N_\mathrm{Ch}^{(2)}$. The measured gate charge is $\gamma_\mathrm{geo}Q_\mathrm{4D\mathrm{-}QHE}$, where $\gamma_\mathrm{geo}=C_\mathrm{total}/C_\mathrm{S}$ is the geometric attenuation factor.}
  \label{fig:figS6}
\end{figure}

\begin{table*}[t]
\centering
\caption{Comparison between the TME and the QAH effect in terms of bulk response, surface charge response, and experimental accessibility.}
\label{tab:comparison_QAH_TME}
\footnotesize
\renewcommand{\arraystretch}{1.3}
\setlength{\tabcolsep}{5pt}

\begin{tabular}{@{}P{2.0cm}P{3.3cm}P{3.5cm}P{3.3cm}P{3.2cm}@{}}
\toprule
& \textbf{Bulk response}
& \textbf{Surface-charge response}
& \textbf{Out-of-plane detection (this work)}
& \textbf{In-plane transport / pumping} \\
\midrule

\textbf{TME}\\
(4D QHE)
&
$\Delta P=\dfrac{e^2}{2h}N_\mathrm{Ch}^{(2)}\Delta B$.
&
Opposite surface charges:
$+\dfrac{e^2}{2h}\Delta B$ on one surface and
$-\dfrac{e^2}{2h}\Delta B$ on the other.
&
$I_{\mathrm{TME}}=i\omega\dfrac{C_{\mathrm{total}}}{C_{\mathrm{S}}}A_\mathrm{S}\dfrac{e^2}{2h} B_\mathrm{AC}$.

Directly accessible to TME polarization.
&
$\sigma_{xy}=0$, $I_{\mathrm{pump}}=0$.

Not accessible to TME polarization.  \\

\addlinespace[0.6em]

\textbf{QAH}\\
(2D QHE)
&
$\Delta P=0$ due to cancellation between
$N_\mathrm{Ch}^{(2)}=+1$ and $-1$ domains.
&
Same-sign surface charges:
$\pm\dfrac{e^2}{2h}\Delta B$ on each surface.

Total charge accumulation:
$\pm\dfrac{e^2}{h}\Delta B$.
&
$I_{\mathrm{QAH}}=i\omega A_\mathrm{S}\dfrac{e^2}{h} B_\mathrm{AC}$.

Accessible to 2D-QHE through charge accumulation.
&
$\sigma_{xy}=\dfrac{e^2}{h}$, $I_{\mathrm{pump}}(r)= -i\omega \sigma_{xy} \pi r^2 B_\mathrm{AC}$.

Accessible to 2D QHE and quantized $\sigma_{xy}$.\\

\bottomrule
\end{tabular}
\end{table*}

In this section, we analyze the out-of-plane charge measurement schemes for the total charge accumulation in the QAH system and for the TME polarization charge. For clarity, we consider the ideal limit without dissipation.

For the single-gate QAH charge measurement, the intrinsic field-induced total charge accumulation is
\begin{equation}
    Q_\mathrm{2D-QHE} = A_\mathrm{S} \eta_0 = A_\mathrm{S} \frac{e^2}{h}\Delta B,
\end{equation}
where $\eta_0 = \frac{e^2}{h}\Delta B$ is the intrinsic field-induced 2D charge accumulation density, as derived in SI Section~\ref{sec:SMsection7}, and $A_\mathrm{S}$ is the sample area. In QAH systems formed by magnetically doped topological insulators, the top and bottom surfaces each contribute a half-quantized surface Hall conductance and therefore each carry one half of the total accumulated charge, which can be regarded as source charges in the effective circuit (Fig.~\ref{fig:figS6}\textbf{a},\textbf{b}). In the single-gate geometry, the top gate is capacitively coupled to the sample, and the sum of the charge accumulations on the two sample surfaces is measured through the grounded gate line by the current preamplifier. As a result, the measured charge is directly
\begin{equation}
    q = Q_\mathrm{2D-QHE}.
\end{equation}

Under an AC magnetic field $B_\mathrm{AC}$ with an angular frequency $\omega$, the corresponding current signal is
\begin{equation}
    I_\mathrm{QAH} = i\omega Q_\mathrm{2D-QHE}
    = i\omega A_\mathrm{S}\frac{e^2}{h} B_\mathrm{AC}.
\end{equation}

For the TME case, the intrinsic polarization produces opposite surface charges $\pm Q_\mathrm{4D-QHE}$ on the two surfaces, with
\begin{equation}
    Q_\mathrm{4D-QHE}
    =
    A_\mathrm{S} \Delta B \cdot \frac{e^2}{2h} N_\mathrm{Ch}^{(2)},
\end{equation}
\cite{SCZhang2001_4DQH,qi2008topological,Vanderbilt2009_3DTI}
which can again be treated as source charges in the effective circuit. Because the total polarization charge on top and bottom surfaces sums to zero, the TME signal cannot be detected in a single-gate geometry and instead requires a dual-gate configuration (Fig.~\ref{fig:figS6}\textbf{c},\textbf{d}). In this case, three geometric capacitances affect the measured signal: the top-gate dielectric capacitance $C_\mathrm{TG}$, the bottom-gate dielectric capacitance $C_\mathrm{BG}$, and the sample geometric capacitance between the top and bottom surfaces $C_\mathrm{S}$.

Denoting $q_1$, $q_2$, and $q_\mathrm{S}$ as the charges on the lower plates (see Fig.~\ref{fig:figS6}\textbf{d}) of $C_\mathrm{TG}$, $C_\mathrm{BG}$, and $C_\mathrm{S}$, respectively, charge conservation at the top and bottom sample surfaces gives
\begin{equation}
    Q_\mathrm{4D-QHE} = q_1 - q_\mathrm{S},
\qquad
- Q_\mathrm{4D-QHE} = q_\mathrm{S} - q_2.
\end{equation}

Since the top and bottom current preamplifiers are both effectively grounded, the total voltage drop across the three capacitors must vanish:
\begin{equation}
\frac{q_1}{C_\mathrm{TG}}+\frac{q_\mathrm{S}}{C_\mathrm{S}}+\frac{q_2}{C_\mathrm{BG}}=0.
\label{eq:SM_TME_voltage_loop}
\end{equation}
Solving these equations yields
\begin{equation}
q_1=q_2=\frac{C_\mathrm{total}}{C_\mathrm{S}}\,Q_\mathrm{4D-QHE}
\equiv \gamma_\mathrm{geo}Q_\mathrm{4D-QHE},
\qquad
C_\mathrm{total}
\equiv
\left(
\frac{1}{C_\mathrm{S}}+\frac{1}{C_\mathrm{TG}}+\frac{1}{C_\mathrm{BG}}
\right)^{-1},
\label{eq:SM_TME_uncomp}
\end{equation}
where the corresponding geometric attenuation factor is
\begin{equation}
\gamma_\mathrm{geo}=\frac{C_\mathrm{total}}{C_\mathrm{S}}.
\label{eq:SM_TME_gamma_uncomp}
\end{equation}

Under $B_\mathrm{AC}$ with an angular frequency $\omega$, the corresponding current signal is
\begin{equation}
    I_\mathrm{TME} = i\omega q_1
    = i\omega \frac{C_\mathrm{total}}{C_\mathrm{S}} A_\mathrm{S}\frac{e^2}{2h} B_\mathrm{AC}.
\end{equation}

Therefore, our out-of-plane charge detection scheme can directly measure the total charge accumulation originating from the QAH effect in a single-gate geometry and the TME polarization in a dual-gate geometry. In contrast, conventional in-plane measurements, such as transport probes of Laughlin charge pumping \cite{1992_Russia_GaAs_chargepumping&Streda, RIKEN_CBST_chargepumping}, cannot access the TME signal because they probe the net sum of the in-plane Hall responses from the top and bottom surfaces, which strictly vanishes in the TME configuration regardless of whether the quantized charge accumulation (with opposite signs) develops on the top and bottom surfaces. Our out-of-plane detection scheme thus provides a unique advantage for accessing quantized topological bulk responses. A detailed comparison of the QAH and TME effects in terms of their bulk and surface state properties, and their relation to different measurement schemes is summarized in Table~\ref{tab:comparison_QAH_TME}.

\section{Analytical derivation of the charge accumulation and Laughlin charge pumping in the Corbino-disk-shaped sample with finite dissipation}
\label{sec:SMsection7}


\begin{figure}[htbp]
  \centering
  \includegraphics[width=1 \linewidth]{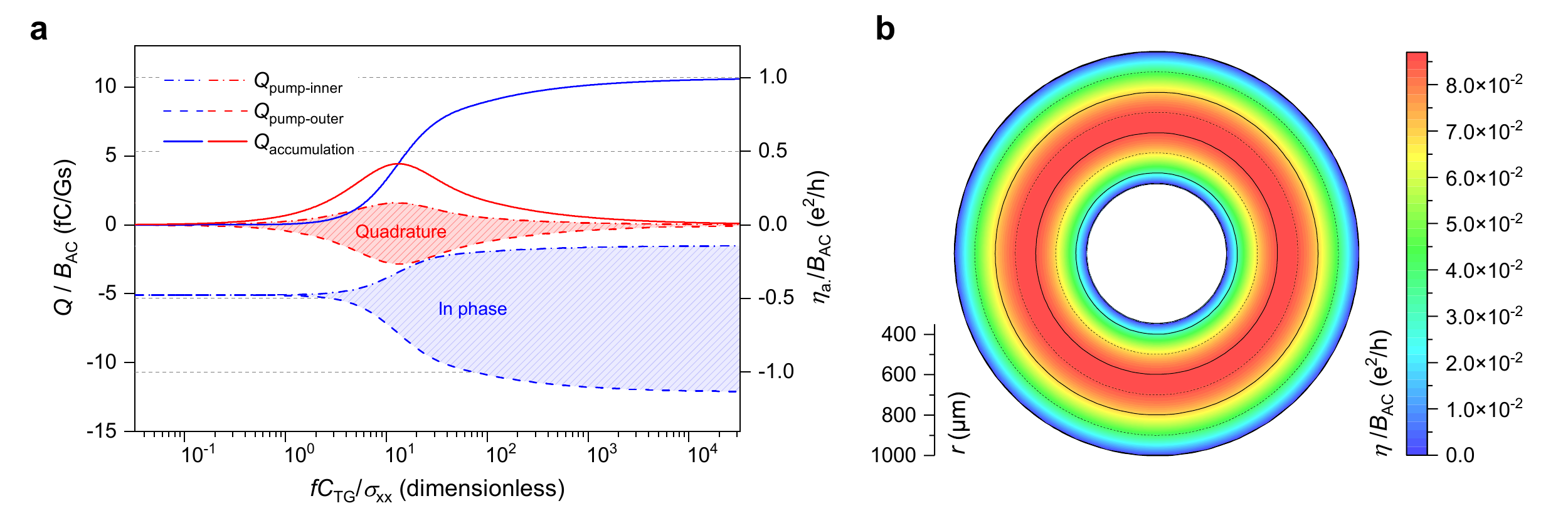}
  \caption{\textbf{Simulation of charge accumulation and charge pumping under analytical dissipation.} \textbf{a}, $f C_\mathrm{TG}/\sigma_{xx}$ dependence of charge accumulation $Q_\mathrm{accumulation}$, charge pumping at inner contact $Q_\mathrm{pump-inner}$ and charge pumping at outer contact $Q_\mathrm{pump-outer}$ (all normalized by $B_\mathrm{AC}$). The simulation uses Eqs. \eqref{accumulated_charge_Corbino} and \eqref{pumped_charge_Corbino} with real sample parameters measured separately. The shaded areas show the difference between $Q_\mathrm{pump-inner}$ and $Q_\mathrm{pump-outer}$, which equals $Q_\mathrm{accumulation}$. The right axis displays the average charge accumulation density $\eta_\mathrm{a.}$ corresponding to $Q_\mathrm{accumulation}$. \textbf{b}, Spatial distribution of normalized charge accumulation density $\eta/B_\mathrm{AC}$ (AC amplitude) at $f = 277.777$ Hz. The simulation uses Eq. \eqref{charge_density_distribution}.}
  \label{fig:figS7}
\end{figure}

To calculate the charge accumulation result including the dissipation effect, we first derive the differential equation for the complex charge density distribution $\eta(\mathbf{r},t)$, and then solve the equation using boundary conditions. Three aspects need to be analyzed, namely, the transport properties of the QAH sample, the relation between the charge accumulation and the in-plane electric field, and the relation between an out-of-plane magnetic field variation and the in-plane electric field. We start from the in-plane transport properties of the QAH thin film

\begin{equation}
    \left[\begin{array}{l}
    J_x \\
    J_y
    \end{array}\right]=\left[\begin{array}{cc}
    \sigma_{x x} & \sigma_{x y} \\
    -\sigma_{x y} & \sigma_{x x}
    \end{array}\right]\left[\begin{array}{l}
    E_x \\
    E_y
    \end{array}\right]
    \label{2D_conductivity}
\end{equation}

where $\sigma_{xx}$ and $\sigma_{xy}$ are the longitudinal and Hall conductances of the QAH sample, respectively, and $\mathbf{J}$ and $\mathbf{E}$ are the in-plane current density and electric field, respectively.

If a time-varying uniform magnetic field $\mathbf{B} = B \mathbf{e}_z$ is applied perpendicular to the sample plane, an in-plane circulating electric field is generated according to Faraday's law

\begin{equation}
    -\frac{\partial \mathbf{B}}{\partial t} = \nabla \times \mathbf{E} \ \rightarrow \frac{\partial B_z}{\partial t} = \frac{\partial E_x}{\partial y} - \frac{\partial E_y}{\partial x}
    \label{Maxwell_fieldchange}
\end{equation}

The induced charge accumulation density $\eta$ is described by the law of charge conservation

\begin{equation}
    \frac{\partial \eta}{\partial t} = - \nabla \cdot \mathbf{J} = - \frac{\partial J_x}{\partial x} - \frac{\partial J_y}{\partial y}
    \label{charge_conservation}
\end{equation}

By plugging Eqs. \eqref{2D_conductivity} and \eqref{Maxwell_fieldchange} into Eq. \eqref{charge_conservation},
\begin{equation}
\begin{split}
    \frac{\partial \eta}{\partial t} &= - \frac{\partial}{\partial x}\left( \sigma_{xx} E_x +\sigma_{xy} E_y \right) - \frac{\partial}{\partial y}\left( \sigma_{xx} E_y -\sigma_{xy} E_x \right) \\
    &= \sigma_{xy} \left( \frac{\partial E_x}{\partial y} - \frac{\partial E_y}{\partial x} \right) - \sigma_{xx} \left( \frac{\partial E_x}{\partial x} + \frac{\partial E_y}{\partial y} \right)\\
    &= \sigma_{xy} \frac{\partial B_z}{\partial t} - \sigma_{xx} \nabla \cdot \mathbf{E}\\
    &= \sigma_{xy} \frac{\partial B_z}{\partial t} + \sigma_{xx} \nabla^2 U,
\end{split}
\label{charge_accumulation}
\end{equation}
where the first term describes the intrinsic charge accumulation, the second term describes the dissipation effect, and $U$ is the electric potential (voltage) in the sample. The ideal charge accumulation density
\begin{equation}
    \eta_0 = \sigma_{xy} \Delta B
\end{equation}
is obtained from Eq. \eqref{charge_accumulation} when dissipation is negligible.

For a sample with a grounded top gate, the accumulated charge density $\eta$ determines the in-plane electric field and resulting dissipation, which can be analyzed from the out-of-plane electric field and electrical potential. Since the length scale of the charge-accumulation distribution is much larger than the thickness of the gate insulator, the electric field (generated by the charge) in the insulator is approximately perpendicular to the sample. As a result, the electrical potential of a point in the sample is exclusively determined by the charge accumulation density of that location,
\begin{equation}
    U = \frac{w}{\epsilon} \eta = \frac{A_\mathrm{TG}}{C_\mathrm{TG}} \eta,
    \label{TG_relation}
\end{equation}
where $w$ and $\epsilon$ are the thickness and dielectric constant of the insulating layer, respectively.

Substituting Eq. \eqref{TG_relation} into Eq. \eqref{charge_accumulation},
\begin{equation}
    \frac{\partial \eta}{\partial t} = \sigma_{xy} \frac{\partial B_z}{\partial t} + \sigma_{xx} \frac{A_\mathrm{TG}}{C_\mathrm{TG}} \nabla^2 \eta
    \label{general_differential_equation}
\end{equation}

which is the general form of the differential equation of the charge accumulation density. The only unknown variable in this equation is the charge density $\eta$, which can be solved with boundary conditions. 

Given a sinusoidal alternating magnetic field $B_z(t) = B_{\mathrm{AC}} e^{i\omega t}$ ($\omega \equiv 2 \pi f$), the response of the charge accumulation also has the time factor $e^{i \omega t}$. Combining amplitude and phase into a complex value $\tilde{\eta}(\mathbf{r})$, we get the complex differential equation of the charge accumulation

\begin{equation}
    (1 + i\frac{ \sigma_{xx} A_\mathrm{TG}}{2 \pi f C_\mathrm{TG}} \nabla^2) \tilde{\eta} = \sigma_{xy} B_{\mathrm{AC}}
    \label{AC_differential_equation}
\end{equation}

For a Corbino-disk-shaped QAH sample with inner ring radius $r_1$ and outer ring radius $r_2$, we conduct calculations in cylindrical coordinates $(r, \theta)$. The boundary conditions of the Corbino disk sample are (grounded contact electrodes)

\begin{equation}
U(r_1) = U(r_2) = 0 \rightarrow \eta(r_1) = \eta(r_2) = 0
\label{boundary_condition}
\end{equation}

where the relation described in Eq. \eqref{TG_relation} is used. Due to the rotational symmetry of the system, the charge accumulation on a Corbino-disk sample is independent of the polar angle $\theta$, so the equation \eqref{AC_differential_equation} can be simplified to a radial equation with Bessel function solutions

\begin{equation}
    \tilde{\eta}(r) - \sigma_{xy} B_{\mathrm{AC}} + \frac{i \sigma_{xx} A_\mathrm{TG}}{2 \pi f C_\mathrm{TG}} \frac{1}{r} \frac{\partial}{\partial r}\left( r \frac{\partial \tilde{\eta}(r)}{\partial r}\right) = 0
    \label{distribution_differential_equation}
\end{equation}

Define the complex wavenumber for Bessel functions

\begin{equation}
    k \equiv \sqrt{\frac{2 \pi f C_\mathrm{TG}}{i \sigma_{xx} A_\mathrm{TG}}} = (1-i) \sqrt{\frac{\pi f C_\mathrm{TG}}{\sigma_{xx} A_\mathrm{TG}}}
    \label{wave_number}
\end{equation}

Then the general solution of Eq. \eqref{distribution_differential_equation} is

\begin{equation}
    \tilde{\eta}(r) = \sigma_{xy} B_{\mathrm{AC}} + \alpha J_0(kr) + \beta Y_0(kr)
    \label{general_solution}
\end{equation}

where $J$ and $Y$ are Bessel functions of the first and second kind, respectively. The subscript $0$ denotes the order zero. Using the boundary conditions in Eq. \eqref{boundary_condition}, we can solve to obtain

\begin{align}
    \alpha &= \sigma_{xy} B_{\mathrm{AC}} \cdot \frac{Y_0(kr_1) - Y_0(kr_2)}{J_0(kr_1)Y_0(kr_2) - J_0(kr_2)Y_0(kr_1)}
    \label{alpha}
    , \\
    \beta &= \sigma_{xy} B_{\mathrm{AC}} \cdot \frac{J_0(kr_2) - J_0(kr_1)}{J_0(kr_1)Y_0(kr_2) - J_0(kr_2)Y_0(kr_1)}
    \label{beta}
\end{align}

Substituting Eqs. \eqref{alpha} and \eqref{beta} into Eq. \eqref{general_solution}, we can calculate the total accumulated charge in the Corbino-disk sample

\begin{equation}
\begin{split}
  \tilde{Q}_\mathrm{a.} = \, &\int_{r_1}^{r_2} 2 \pi r \, dr \, \tilde{\eta}(r) \\
  = \, &\sigma_{xy} B_{\mathrm{AC}} \Big\{ \pi (r_2^2 - r_1^2) + \\
  &\frac{2\pi}{k \left(J_0(kr_1)Y_0(kr_2) - J_0(kr_2)Y_0(kr_1)\right)} \times \\
  &\left[(Y_0(kr_1) - Y_0(kr_2)) \cdot (r_2 J_1(kr_2) - r_1 J_1(kr_1)) + \right. \\
  &\left. (J_0(kr_2) - J_0(kr_1)) \cdot (r_2 Y_1(kr_2) - r_1 Y_1(kr_1)) \right] \Big\}
\end{split}
\label{accumulated_charge_Corbino}  
\end{equation}

where $k$ is defined in Eq. \eqref{wave_number}. The dissipation factor $g(\sigma_{xx}/fC_{\mathrm{TG}})$ can be extracted as

\begin{equation}
\begin{split}
  g(\sigma_{xx}/fC_{\mathrm{TG}}) = \, &\frac{\tilde{Q}_\mathrm{a.}}{\pi (r_2^2 - r_1^2)\sigma_{xy} B_{\mathrm{AC}}} \\
  = \, &\Big\{ 1 + \frac{2}{k (r_2^2 - r_1^2) \left(J_0(kr_1)Y_0(kr_2) - J_0(kr_2)Y_0(kr_1)\right)} \times \\
  &[(Y_0(kr_1) - Y_0(kr_2)) \cdot (r_2 J_1(kr_2) - r_1 J_1(kr_1)) + \\
  &(J_0(kr_2) - J_0(kr_1)) \cdot (r_2 Y_1(kr_2) - r_1 Y_1(kr_1)) ] \Big\},
\end{split}
\label{dissipation_factor_charge_Corbino}  
\end{equation}
where the ratio $(\sigma_{xx}/fC_\mathrm{TG})$ is described in $k$.

A simulation of the charge accumulation $\tilde{Q}_\mathrm{a.}$ against the dissipation parameter $f C_\mathrm{TG}/\sigma_{xx}$, calculated from Eq. \eqref{accumulated_charge_Corbino}, is plotted in Fig.~\ref{fig:figS7}\textbf{a}, where real geometric parameters of the sample are used ($r_1 = 350$ µm, $r_2 = 1000$ µm, $A_\mathrm{TG}=3.52\times 10^{-6}$ m$^2$). When $f C_\mathrm{TG}/\sigma_{xx}$ increases from 0 to sufficiently large values, the in-phase component of $\tilde{Q}$ increases from $0$ to the intrinsic value of $\pi \sigma_{xy} B_{\mathrm{AC}} (r_2^2 - r_1^2)$, while the quadrature component of $\tilde{Q}$ shows a single peak structure with a value of 0 at two sides.

For a deeper understanding of the dissipation model, especially the effect of the grounded boundary condition, we calculate the charge density distribution in the Corbino disk sample from Eqs. \eqref{general_solution}, \eqref{alpha} and \eqref{beta}, as

\begin{equation}
    \tilde{\eta}(r) = \sigma_{xy} B_{\mathrm{AC}} \left[ 1 + \frac{J_0(kr)(Y_0(kr_1) - Y_0(kr_2)) + Y_0(kr)(J_0(kr_2) - J_0(kr_1))}{J_0(kr_1)Y_0(kr_2) - J_0(kr_2)Y_0(kr_1)} \right]
\label{charge_density_distribution}
\end{equation}

The charge density distribution (AC amplitude), simulated by Eq. \eqref{charge_density_distribution}, is displayed in Fig.~\ref{fig:figS7}\textbf{b}, with the color representing the charge density. $f = 277.777$ Hz and $\sigma_{xx} = 6.2 \times 10^{-7}$ S are used in the simulation. The charge density shows a prominent accumulation in the interior region of the sample, while it decays to zero at both edges, which is consistent with the qualitative model.

The Laughlin charge-pumping value on the Corbino disk sample can also be obtained analytically under finite dissipation. Since the charge pumping and charge accumulation can occur simultaneously in the same state of a sample, we can calculate the charge pumping value following the previous analysis of charge accumulation. The total pumping current $I_\mathrm{pump}$ at a ring of radius $r$ (flowing outward) on the sample is

\begin{equation}
\begin{split}
    I_\mathrm{pump}(r) &= \int_0^{2 \pi} J_r(r, \theta) r \, d\theta = \int_0^{2 \pi} (\sigma_{xx} E_r(r, \theta) + \sigma_{xy} E_{\theta}(r, \theta)) r \, d\theta\\
    &= - 2 \pi r \sigma_{xx} \frac{\partial U(r)}{\partial r} - \sigma_{xy} \pi r^2 \frac{\partial B_z}{\partial t}\\ 
    &= - 2 \pi r \sigma_{xx} \frac{A_\mathrm{TG}}{C_\mathrm{TG}} \frac{\partial \eta}{\partial r} - \sigma_{xy} \pi r^2 \frac{\partial B_z}{\partial t}
\end{split}
\end{equation}

where Eq. \eqref{TG_relation} is used. Considering the complex amplitude $\tilde{I_{r}}$ of $I_r$ in the AC field and using the expression of $k$ and $\tilde{\eta}(r)$ described in Eq. \eqref{wave_number} and Eq. \eqref{charge_density_distribution}, respectively,

\begin{equation}
\begin{split}
    \tilde{I}_\mathrm{pump}(r) &= - 2 \pi r \sigma_{xx} \frac{A_\mathrm{TG}}{C_\mathrm{TG}} \frac{\partial \tilde{\eta}}{\partial r} - i \omega \sigma_{xy} \pi r^2 B_\mathrm{AC}\\
    &= - i \omega \sigma_{xy} B_\mathrm{AC} \left[\pi r^2 + \frac{2\pi r}{k} \frac{J_1(kr)(Y_0(kr_1) - Y_0(kr_2)) + Y_1(kr)(J_0(kr_2) - J_0(kr_1))}{J_0(kr_1)Y_0(kr_2) - J_0(kr_2)Y_0(kr_1)} \right]
\end{split}
\end{equation}

which corresponds to a pumped charge of

\begin{equation}
    \tilde{Q}_\mathrm{pump}(r) = -\sigma_{xy} B_\mathrm{AC} \left[\pi r^2 + \frac{2\pi r}{k} \frac{J_1(kr)(Y_0(kr_1) - Y_0(kr_2)) + Y_1(kr)(J_0(kr_2) - J_0(kr_1))}{J_0(kr_1)Y_0(kr_2) - J_0(kr_2)Y_0(kr_1)} \right]
\label{pumped_charge_Corbino}
\end{equation}

To compare with the charge accumulation, simulations of the charge pumping at both contacts ($\tilde{Q}_\mathrm{pump}(r_1)$ and $\tilde{Q}_\mathrm{pump}(r_2)$) against the dissipation parameter $f C_\mathrm{TG}/\sigma_{xx}$ are also plotted in Fig.~\ref{fig:figS7}\textbf{a}, calculated from Eq. \eqref{pumped_charge_Corbino}. The same parameters as for charge accumulation are used in the simulation. When $f C_\mathrm{TG}/\sigma_{xx}$ is low (severe dissipation), the in-phase components of $\tilde{Q}_\mathrm{pump}(r_1)$ and $\tilde{Q}_\mathrm{pump}(r_2)$ show a consistent value of $\frac{\pi (r_2^2-r_1^2)}{2 \ln{(r_2/r_1)}} \sigma_{xy} B_\mathrm{AC}$, while the quadrature components are nearly zero. The increase of $f C_\mathrm{TG}/\sigma_{xx}$ (suppression of dissipation) induces a difference between $\tilde{Q}_\mathrm{pump}(r_1)$ and $\tilde{Q}_\mathrm{pump}(r_2)$, where the quadrature components undergo opposite single-peak evolutions and finally vanish to zero, while the in-phase components gradually widen to stable values of $\tilde{Q}_\mathrm{pump}(r_1) = \pi r_1^2 \sigma_{xy} B_\mathrm{AC}$ and $\tilde{Q}_\mathrm{pump}(r_2) = \pi r_2^2 \sigma_{xy} B_\mathrm{AC}$. 

The charge conservation can be verified from Eqs. \eqref{accumulated_charge_Corbino} and \eqref{pumped_charge_Corbino}, as follows:

\begin{equation}
    \tilde{Q}_\mathrm{pump}(r_1) - \tilde{Q}_\mathrm{pump}(r_2) = \tilde{Q}_\mathrm{a.} 
\end{equation}

\section{$C_\mathrm{TG}$ measurement of the Corbino disk sample}
\label{sec:SMsection8}

The capacitance $C_\mathrm{TG}$ between the top gate and the sample is a crucial parameter to determine the dissipation factor. We measured the AC impedance between the top gate and contact electrodes to extract this capacitance in situ. For a simple circuit with a capacitor of $C$ and a resistor of $R$ in series, the total impedance (complex value) at frequency $f$ is

\begin{equation}
    Z = R + \frac{1}{i 2 \pi f C}
\end{equation}

When $f \ll 1/(R C)$, the capacitance dominates the impedance with an imaginary value. The effective circuit between the top gate and contact electrodes is a complex mixed result of capacitance and sample resistance. However, when the frequency is small enough, the impedance of this circuit approaches the impedance of a single $C_\mathrm{TG}$. Quantitatively,

\begin{equation}
    1/(i 2 \pi f Z_\mathrm{TG-to-contacts}) \xrightarrow{f \to 0} C_\mathrm{TG}
\end{equation}

\begin{figure}[htbp]
  \centering
  \includegraphics[width=0.8\linewidth]{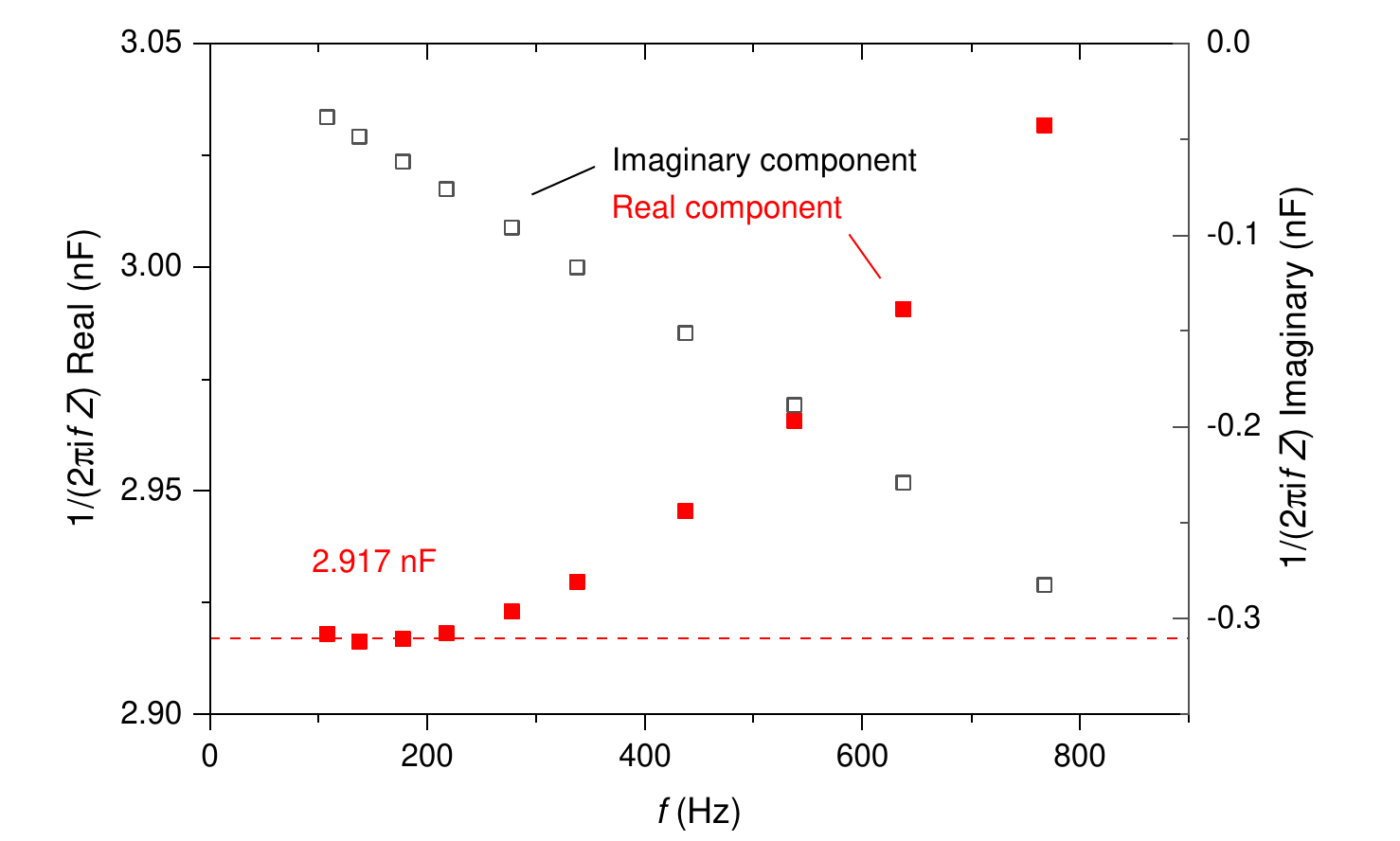}
  \caption{\textbf{Processed impedance between top gate and contact electrodes under 0.5T magnetic field.}}
  \label{fig:figS8}
\end{figure}

We conducted the impedance measurement when the sample was held under the same conditions as in the charge accumulation measurements. An AC voltage of 50  V at various frequencies was applied to the top gate of the Corbino disk sample through the bias circuit of a current preamplifier, while the contact electrodes were AC grounded (with DC gating voltage). The impedance was calculated from the measured current signal and source voltage. The processed data are shown in Fig.~\ref{fig:figS8}, where the real and imaginary parts are plotted separately. At the low frequency limit, the real part of the data approaches a constant value of about 2.917 nF, which is the target capacitance, while the imaginary part approaches zero.

However, the parasitic capacitance of the wire connecting the top gate was included in the result. We measured this parasitic capacitance by disconnecting the sample and remeasuring the wire's AC impedance with the same method; a value of 0.92 nF was obtained. Finally, we obtained the calibrated value of $C_\mathrm{TG}$ as 2.00 nF.

The covered area of the top gate is 3.52 $\mathrm{mm^2}$. Using the relative dielectric constant of approximately 9 for AlO$_x$, it can be calculated that the thickness of the AlO$_x$ layer is 140 nm, which is consistent with the setting parameter of the atomic layer deposition (ALD).

\section{$\sigma_{xx}$ and $C_\mathrm{TG}$ measurements of the simple disk samples}
\label{sec:SMsection9}

\begin{figure}[htbp]
  \centering
  \includegraphics[width=1\linewidth]{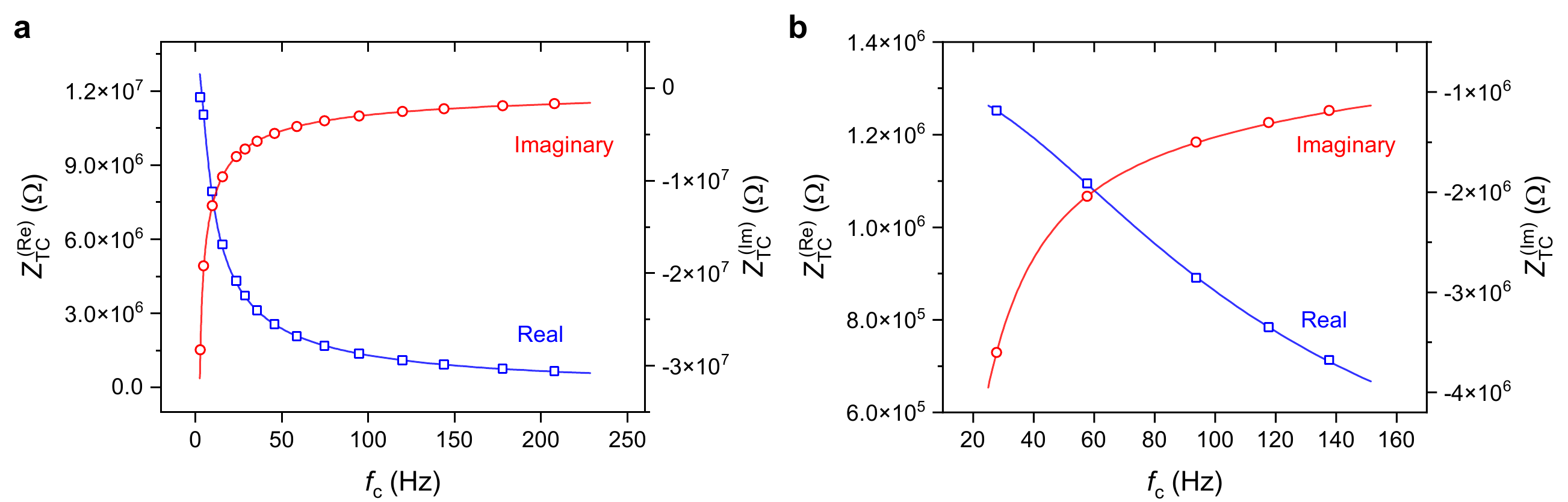}
  \caption{\textbf{$\sigma_{xx}$ and $C_\mathrm{TG}$ fits from the gate-to-contact impedance of simple disk samples.} Frequency dependence of the real and imaginary parts of the gate-to-contact impedance measured on disk samples. Symbols are experimental data and solid lines are fits to the impedance model. 
  \textbf{a}, Data and fits of disk sample I. The sample temperature (approximately 20~mK) and measurement field (1.5~T) correspond to dataset 1 in main text Table~2. 
  \textbf{b}, Data and fits of disk sample II. The sample temperature (approximately 60~mK) and measurement field (0.5~T) correspond to dataset 2 in main text Table~2.}
  \label{fig:figS9}
\end{figure}

\begin{table}[htbp]
\centering
\caption{Fitted $\sigma_{xx}$, $C_\mathrm{TG}$, and $C_\mathrm{p1}$ values for the simple disk samples. Uncertainties reflect one standard deviation from the fit.}
\label{tab:simple_disk_params}
\begin{tabular*}{0.90\linewidth}{@{\extracolsep{\fill}}lccc@{}}
\toprule
Sample & $\sigma_{xx}$ ($10^{-9}$ S) & $C_\mathrm{TG}$ (nF) & $C_\mathrm{p1}$ (nF) \\
\midrule
Disk sample I  & $2.34 \pm 0.04$  & $2.00 \pm 0.03$  & $0.239 \pm 0.003$ \\
Disk sample II & $23.51 \pm 0.07$ & $1.506 \pm 0.002$ & $0.202 \pm 0.002$ \\
\bottomrule
\end{tabular*}
\end{table}

Since the simple disk samples have only a single contact electrode, $\sigma_{xx}$ cannot be directly measured via contacts as in Corbino disk samples. To calibrate $\sigma_{xx}$ and $C_\mathrm{TG}$, we measured the total impedance between the top gate and the contact, $Z_\mathrm{TC}$, as a function of frequency ($f$), and performed a three-parameter fit. $\sigma_{xx}$, $C_\mathrm{TG}$, and $C_\mathrm{p1}$ are fitting parameters, where $C_\mathrm{p1}$ represents the geometric parasitic capacitance between the top gate and the contact.

The analytical expression of $Z_\mathrm{TC}$ for the simple disk samples has been analyzed in another work~\cite{li2026}. Given the fixed sample geometry with $r_0 = 1200$ µm, $Z_\mathrm{TC}$ depends only on the three fitting parameters and $f$:
\begin{equation}
Z_\mathrm{TC}(f, \sigma_{xx}, C_\mathrm{TG}, C_\mathrm{p1})
=
\frac{1}{2 \pi i f}
\left[
C_\mathrm{p1}+C_\mathrm{TG}\frac{2J_1(kr_0)}{kr_0J_0(kr_0)}
\right]^{-1},
\label{eq:TG_to_contact_impedance}
\end{equation}
where $k = k(f, \sigma_{xx}, C_\mathrm{TG})$ is defined in Eq.~\eqref{wave_number}.

For the two disk samples, the $Z_\mathrm{TC}$ measurements were performed under the same temperature and magnetic field conditions as the corresponding datasets in main text Table~2. An AC magnetic field identical to that used for charge accumulation measurements was applied to reproduce the measurement temperature (the AC field frequency is different from the impedance measurement frequency to avoid interference). The measured frequency-dependent real and imaginary components of $Z_\mathrm{TC}$, together with the fits, are shown in Fig.~\ref{fig:figS9}.

The extracted fitting parameters are summarized in Table~\ref{tab:simple_disk_params}. The fitting uncertainties were obtained from the covariance matrix of the least-squares fit in log-parameter space, propagated to the physical parameters. The difference in $C_\mathrm{TG}$ between the two disk samples arises from the different thicknesses of the fabricated top-gate dielectric, whereas the difference in $\sigma_{xx}$ originates from the different measurement temperatures. The resulting difference in parameter scales enables the two $f$-dependence measurements of the charge signal to span a broad range of the combined parameter $f C_\mathrm{TG}/\sigma_{xx}$ in main text Fig.~4.

\section{Calibration of the AC magnetic field through a Hall sensor}
\label{sec:SMsection10}

The charge measurements require a precise AC magnetic field, which was generated by driving the superconducting coil with an AC source voltage. However, due to the coil's inductance and the generation of eddy currents in nearby metallic structures, the relationship between the source voltage and the resulting magnetic field is complex and strongly frequency-dependent. This precludes a simple analytical calculation and necessitates a direct, in-situ calibration. 

\begin{figure}[htbp]
  \centering
  \includegraphics[width=1\linewidth]{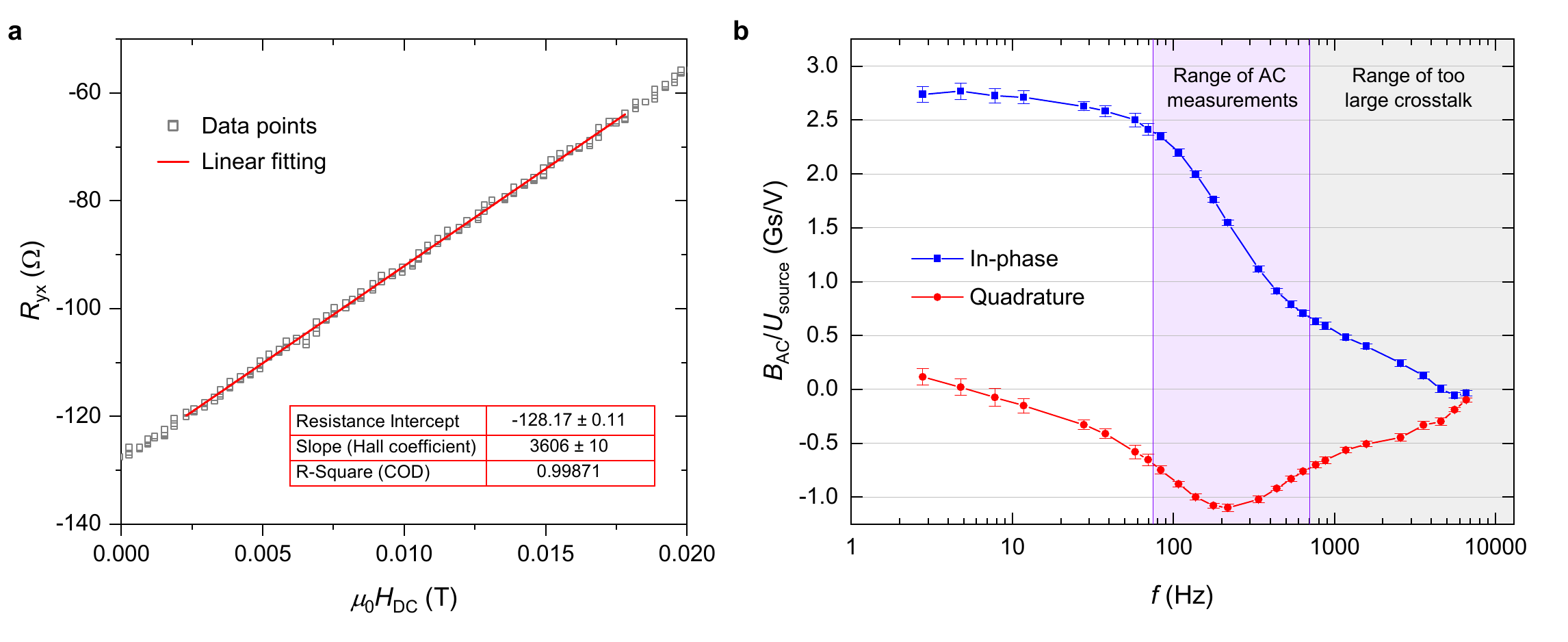}
  \caption{\textbf{The calibration of the AC magnetic field via a Hall sensor.} \textbf{a}, The linear fitting of the sensor's Hall coefficient in the DC field range of 0.002 T to 0.018 T. \textbf{b}, The frequency dependence of the AC field per unit of source voltage. The error bars on data points represent standard errors calculated from repeated measurements.}
  \label{fig:figS10}
\end{figure}

We used a Hall sensor placed near the sample to calibrate the AC field at various frequencies. A constant DC current was applied to the sensor at a DC field of 0.01 T, while the resulting AC Hall voltage was measured with a lock-in amplifier referenced to the source voltage. The AC magnetic field ($B_\mathrm{AC}$) was then calculated from this voltage using the sensor's pre-calibrated Hall coefficient ($k_\mathrm{Hall} = \Delta \rho_{yx}/\Delta B$)

\begin{equation}
    B_\mathrm{AC} = U_\mathrm{Hall}/I_\mathrm{sensor}/k_\mathrm{Hall}
\end{equation}

The Hall coefficient of $(3606 \pm 10)$ $\Omega/\mathrm{T}$ was determined from the linear fitting of the Hall resistance around a DC field of 0.01 T (Fig.~\ref{fig:figS10}\textbf{a}). Fig.~\ref{fig:figS10}\textbf{b} presents the resulting frequency dependence of the calibrated AC field per unit of source voltage. The in-phase and quadrature components of the field relative to the source voltage are plotted separately, showing a smooth dependence on frequency. The middle frequency region with relatively high frequencies but small crosstalk was used for charge signal measurements. For the final AC charge data analysis, the raw charge signal—which was measured using the coil's source voltage as the lock-in reference—was normalized at each frequency using this calibration to yield the target signal per unit of AC magnetic field.

\section{The effect of chemical potential variation (quantum capacitance) on the charge accumulation model}

In the derivations and analyses above, we have neglected the effect of chemical potential variation on charge transport. The charge accumulation corresponds to a variation in electron density and thus causes the Fermi level to shift, as
\begin{equation}
    d \mu = -\frac{1}{e}\frac{\partial \mu}{\partial n} d \eta.
    \label{chemical_potential_diff}
\end{equation}
Here $n$ is the electron density, $e$ is the absolute value of the electron charge, and we have used $\eta=-en$ for the accumulated charge density. If the charge response associated with the chemical potential is further included, an additional diffusion current driven by the chemical potential gradient should be introduced as
\begin{equation}
    \mathbf{J}_{\mu} = \frac{\sigma_{xx}}{e} \nabla \mu .
\label{chemical_potential_current}
\end{equation}
Correspondingly, Eq.~\eqref{charge_accumulation} is modified to
\begin{equation}
    \frac{\partial \eta}{\partial t} = \sigma_{xy} \frac{\partial B_z}{\partial t} - \sigma_{xx} \nabla \cdot \mathbf{E} - \frac{\sigma_{xx}}{e} \nabla^2 \mu .
\label{charge_accumulation_chemical_potential}
\end{equation}

Using the relation between the electrostatic potential and the accumulated charge density introduced in Eq.~\eqref{TG_relation}, Eq.~\eqref{general_differential_equation} is then modified as
\begin{equation}
\begin{split}
    \frac{\partial \eta}{\partial t} &= \sigma_{xy} \frac{\partial B_z}{\partial t} + \sigma_{xx} \frac{A_\mathrm{TG}}{C_\mathrm{TG}} \nabla^2 \eta - \frac{\sigma_{xx}}{e} \nabla^2 \mu\\
    &= \sigma_{xy} \frac{\partial B_z}{\partial t} + \sigma_{xx} \frac{A_\mathrm{TG}}{C_\mathrm{TG}} \nabla^2 \eta + \frac{\partial \mu}{\partial n}\frac{\sigma_{xx}}{e^2} \nabla^2 \eta\\
    &= \sigma_{xy} \frac{\partial B_z}{\partial t} + \sigma_{xx} A_\mathrm{TG} \left(\frac{1}{C_\mathrm{TG}} + \frac{1}{C_\mathrm{q}}\right) \nabla^2 \eta .
\end{split}
    \label{differential_equation_chemical_potential}
\end{equation}
Here $C_\mathrm{q}$ is the quantum capacitance of the system, defined as
\begin{equation}
    C_\mathrm{q} = A_\mathrm{TG} e^2 \frac{\partial n}{\partial \mu}.
\end{equation}

Therefore, the influence of chemical potential variation on the charge response can be absorbed into a correction to the gate capacitance through the quantum capacitance. Equivalently, $C_\mathrm{TG}$ should be replaced by the total capacitance
\begin{equation}
    C_\mathrm{TG}^\mathrm{(all)} = \left(\frac{1}{C_\mathrm{TG}}+\frac{1}{C_\mathrm{q}}\right)^{-1}.
\end{equation}
After this replacement, the form of the transport equation remains unchanged. Although direct quantum capacitance data for the present CBST samples are not available, existing quantum capacitance measurements on other topological insulator systems indicate that the minimum quantum capacitance per unit area is typically on the order of $10~\mathrm{mF/m^2}$, which is much larger than the geometric capacitance per unit area in our experiment, approximately $0.4~\mathrm{mF/m^2}$~\cite{inhofer2017, inhofer2018, dartiailh2020}. Thus, the correction from the quantum capacitance is expected to be small. Moreover, in our capacitance calibration experiment (Section~\ref{sec:SMsection9}), the sample is in the same state as that used for the charge accumulation measurements. Therefore, the calibrated capacitance already includes the contribution from the quantum capacitance, and the directly calibrated value corresponds to $C_\mathrm{TG}^\mathrm{(all)}$.

In summary, all theoretical derivations and comparisons with the experimental data in this work remain valid when the variation of the chemical potential is taken into account.

\section{Distinction between charge accumulation measurements in the current manuscript and the Streda formula for charge accumulation}

In this section, we briefly discuss the distinction between our frequency-dependent charge-accumulation results (non-equilibrium state) and the Streda formula \cite{streda1982} (equilibrium state). Both describe a quantized relationship between charge accumulation and magnetic field; however, their regimes of applicability and underlying assumptions are fundamentally different.

In general, our model describes a transport response generated by the electric field induced through time-dependent AC magnetic fields. This response depends solely on the Hall and longitudinal conductance of the system and does not require assumptions about equilibrium electron distributions or an ideal energy gap, as assumed in the Streda formula. Therefore, our model is applicable to a broader range of realistic samples and varying magnetic-field conditions. 

Crucially, the field-induced charge accumulation in our model generates a spatially non-uniform chemical potential distribution, reflecting an intrinsically non-equilibrium state that violates the key condition required for the Streda formula. This non-equilibrium character persists even in the low-frequency limit, since, using Eq. \eqref{chemical_potential_diff}, the chemical potential
\begin{equation}
    \mu = \mu_0 - \frac{1}{e} \frac{\partial \mu}{\partial n} \eta,
    \label{chemical_potential}
\end{equation}
depends linearly on the induced charge density $\eta$, and $\eta$ is always spatially inhomogeneous in our model ($\eta = 0$ for the sample edge, as it is grounded, while the interior region of the sample has finite $\eta$ induced by the applied AC magnetic field; thus, $\eta$ cannot be homogeneous). Consequently, using Eqs. \eqref{TG_relation} and \eqref{chemical_potential}, the electrochemical potential
\begin{equation}
    \mu^* = \mu - eU = \mu_0 - \left( \frac{1}{e} \frac{\partial \mu}{\partial n} + \frac{ew}{\epsilon} \right) \eta,
\end{equation}
is also inhomogeneous and has a generally finite gradient. When combined with a finite $\sigma_{xx}$ in a real sample, this gradient provides current channels for charge density relaxation, constituting precisely the dissipative response captured by our model. The frequency dependence is corroborated by the experimental data shown in Fig.~4\textbf{b},\textbf{c} of the main text. The situation addressed by our model is therefore fundamentally different from that described by the Streda formula, which pertains strictly to the static equilibrium state.

\bibliographystyle{sn-nature}

\bibliography{Supplementary_info}